\documentclass[11pt,a4paper,epsf,epsfig,psfrag]{article}
\usepackage{jheppub}
\usepackage{amsmath,graphicx,amsfonts, mathrsfs,amssymb}
\usepackage{amsmath,epsfig}
\usepackage{multirow,longtable,rotating}
\newbox\pippobox
\usepackage{amssymb,amsfonts}
\usepackage{latexsym}
\usepackage{epsfig}
\usepackage{pdfsync}

\def\be{\begin{equation}}
\def\ee{\end{equation}}
\def\bea{\begin{eqnarray}}
\def\eea{\end{eqnarray}}

\newcommand{\beq}{\begin{equation}}
\newcommand{\eeq}{\end{equation}}
\newcommand{\beqa}{\begin{eqnarray}}
\newcommand{\eeqa}{\end{eqnarray}}
\newcommand{\beqar}{\begin{eqnarray*}}
\newcommand{\eeqar}{\end{eqnarray*}}

\renewcommand{\eqref}[1]{(\ref{#1})}

\catcode`\@=12

\title{Gravitation waves from QCD and electroweak phase transitions}
\author{Yidian Chen$^{a,b}$,Mei Huang$^{a,b,c}$,Qi-Shu Yan$^{b}$ \\
$^{a}$ {Institute of High Energy Physics, Chinese Academy of Sciences, Beijing, China} \\
$^{b}$ {School of Physics Sciences, University of Chinese Academy of Sciences, Beijing 100039, China}\\
$^{c}$ {Theoretical Physics Center for Science Facilities, Chinese Academy of Sciences, Beijing, China} \\
}

\abstract{We investigate the gravitation waves produced from QCD and electroweak phase transitions in the early universe by using a 5-dimension holographic QCD model and a holographic technicolor model. The dynamical holographic QCD model is to describe the pure gluon system, where a first order confinement-deconfinement phase transition can happen at the critical temperature around 250 ${\rm MeV}$. The minimal holographic technicolor model is introduced to model the strong dynamics of electroweak, it can give a first order electroweak phase transition at the critical temperature around 100-360 ${\rm GeV}$. We find that for both GW signals produced from QCD and EW phase transitions, in the peak frequency region, the dominant contribution comes from the sound waves, while away from the peak frequency region the contribution from the bubble collision is dominant. The peak frequency of gravitation wave determined by the QCD phase transition is located around $10^{-7}$ Hz which is within the detectability of FAST and SKA, and the peak frequency of gravitational wave predicted by EW phase transition is located at  $0.002-0.007$ Hz, which might be detectable by BBO, DECIGO, LISA and ELISA.}


\begin{document}
\maketitle

\section{Introduction}

Predicted by Albert Einstein on the basis of general relativity, gravitational radiation is generated by the changes of the curvature of spacetime and propagates outwards as a wave at the speed of light \cite{Einstein:1916cc, Einstein:1918btx}. On February 11, 2016, the LIGO and Virgo Scientific Collaboration \cite{Abbott:2016blz} announced the first observed gravitational waves (GWs) signal in the detectors of LIGO. The gravitational waves, named GW150914, were originated from a binary black hole merger. Recently, the gravitational waves, named GW170817, originated from a binary neutron star inspiral were observed in LIGO \cite{TheLIGOScientific:2017qsa}. The Nobel Prize in Physics 2017 was awarded to Rainer Weiss, Kip Thorne and Barry Barish for their decisive contributions to the LIGO detector and the observation of gravitational waves, which opens a new exciting era for astronomy and cosmology.

The GWs can be roughly divided into three categories \cite{Cai:2017cbj}: 1) GWs can be produced through various astrophysical processes, such as compact binary inspirals, explosion of supernova and spherically asymmetric spinning neutron stars, among them, binaries systems are the main sources for detecting GWs through ground-based detectors LIGO, Virgo and Space-based interferometers LISA, DECIGO, BBO. 2) The primordial GWs can be produced in the very early stages of the Universe, such as cosmic strings, the inflation and reheating epochs, and these primordial GWs have unique imprint on the cosmic microwave background; 3) GWs can be produced from cosmological phase transitions in the early universe, such as GUT, electroweak (EW) and Quantum chromodynamics (QCD)  phase transition, and these GWs can tell us the evolution of the universe.

During the evolution of the universe, several phase transitions might have occurred. If it is of a first order phase transition, it can generate gravitational waves. When two local minima of a free energy coexist in a certain temperature range, the scalar field can enter into the broken phase from the symmetric phase via quantum tunnelling or thermally fluctuation, which can lead the nucleation of bubbles in the metastable sea. If this process  is fast enough when compared to the rate of expansion Hubble parameter $H$, the bubbles will expand and collide with each other and produce the GWs \cite{Kosowsky:1992rz,Kosowsky:1992vn,Caprini:2015zlo,Jinno:2015doa}.

When the universe cools down to around several hundreds ${\rm GeV}$, the EW phase transition could happen. Although the phase transition in the electroweak sector of the Standard Model is crossover \cite{Kajantie:1996mn,Gurtler:1997hr,Csikor:1998eu}, first order phase transition is predicted in many extended scenarios beyond the Standard Model \cite{Barger:2007im,Profumo:2007wc,Damgaard:2015con,Vaskonen:2016yiu,Beniwal:2017eik,Chen:2017qcz,Cline:1996mga,Fromme:2006cm,Dorsch:2013wja,Haarr:2016qzq,Gunion:1989ci,FileviezPerez:2008bj}. Therefore, the GW signals from the EW phase transition, which could be detected by LISA \cite{Caprini:2015zlo}, can shed some lights on the new physics beyond the standard model. Moreover, the first order phase transition is also favoured in order to produce the observed baryon asymmetry \cite{Kuzmin:1985mm,Shaposhnikov:1987tw} via the electroweak baryogenesis mechanism. Therefore, gravitational waves physics has opened a new window for the research of the fundamental laws of particle physics and cosmology. In particular, it can serve as an important tool to study the dynamics origin of the EW symmetry breaking and new physics beyond the standard model. As one of the solutions to tackle the hierarchy problem, technicolor models was first introduced by Weinberg \cite{Weinberg:1975gm} and Susskind \cite{Susskind:1978ms}. A more realistic technicolor model, like the walking technicolor \cite{Yamawaki:1985zg,Bando:1986bg,Bando:1987we} scenario, predicts a composite scalar boson techni-dilaton (TD) for a candidate of 125 GeV boson. Similar to strong interaction of the QCD, the walking technicolor model can predict the EW phase transition and GWs generation in the evolution of the universe \cite{Cline:2008hr,Jarvinen:2009mh}.

When the universe further cools down to around several hundreds ${\rm MeV}$, the QCD phase transition happens, and the chiral symmetry is spontaneously broken and color degrees of freedom is confined. Exploring the QCD phase structure under extreme conditions is one the most important tasks for heavy ion collisions, especially in the Relativistic Heavy Ion Collisions (RHIC) and Large Hadron Collider (LHC), where two accelerated nucleus with relativistic velocities collide to create the hot quark-gluon plasma, which is normally called "little bang". Lattice QCD calculation shows that  the phase transition for three light flavors is of smooth crossover at small baryon chemical potentia and high temperature \cite{Fodor:2001au,Ding:2015ona}, and for heavy and static quarks or pure gauge theory,  the QCD phase transition is of first order \cite{Lucini:2012wq}. If the QCD phase transition is flavor dependent and happens sequentially \cite{Xu:2011pz}, there might be chances for the appearance of first order QCD phase transition in the early universe. The GWs detection offers one more experimental tool to explore the QCD phase structure.

In order to tackle strongly coupled gauge theories (see \cite{Aharony:1999ti,Aharony:2002up,Zaffaroni:2005ty,Erdmenger:2007cm} for review), the anti-de Sitter/conformal field theory (AdS/CFT) correspondence \cite{Maldacena:1997re,Gubser:1998bc,Witten:1998qj} or general gravity/gauge duality was proposed. In recent decades, many properties of QCD, for example, meson spectra \cite{Erlich:2005qh,DaRold:2005vr,Karch:2006pv,mesons}, phase transitions and baryon number susceptibilities \cite{bns} have been investigated from both top-down and bottom-up models. Furthermore, new extensions beyond the standard models, such as technicolor models \cite{Haba:2010hu,Matsuzaki:2012xx,htc} and composite higgs models \cite{hchm} have also been studied in the context of AdS/CFT.

In this work, by using a 5D dynamical holographic QCD model and a holographic technicolor model we investigate the first order phase transitions and predict the GW signals of these two models. The paper is organized as following: In Sec.\ref{sec:setup-qcd} we introduce the dynamical holographic QCD model for pure gluon system, and describe the first order deconfinement phase transition. Similar to holographic QCD model, in Sec.\ref{sec:setup-tc}we construct the five dimensional holographic technicolor model, where a first order phase transition for EW can be described. We calculate the GWs from EW and QCD phase transitions in Sec.\ref{sec:gw}. Finally, a short summary is given in Sec.\ref{sec:sum}.

\section{First order confinement-deconfinement phase transition}
\label{sec:setup-qcd}

\subsection{Quenched dynamical holographic QCD model}

In order to tackle the challenging from strong coupling in the infrared (IR) of QCD,  in recent decades, the anti-de Sitter/conformal field theory correspondence or the gauge/gravity duality \cite{Maldacena:1997re,Gubser:1998bc,Witten:1998qj} has been widely applied in investigating hadron physics,
strongly coupled quark gluon plasma, QCD phase transitions and transport properties. It can be regarded as an general principle that for any d-dimensional quantum field theory (QFT) there exists a dual theory of quantum gravity living in (d + 1)-dimensions, and the gravitational description becomes classical when the QFT is strongly-coupled. Here the extra dimension, i.e., the 5th-dimension can be also interpreted as an energy scale or renormalization group (RG) flow in the QFT \cite{Adams:2012th}.

In the past decade, much effort has been paid from both top-down and bottom-up methods on constructing a realistic holographic QCD model. From bottom-up, the most  economic way of breaking the conformal symmetry  is to add a proper deformed warp factor in front of the ${\rm AdS}_5$ metric, which can capture the main non-perturbative QCD features. For example, a quadratic correction in front of the warp factor of ${\rm AdS}_5$ geometry \cite{Andreev:2006ct} or a deformed warp factor which mimics the QCD running coupling \cite{Pirner:2009gr} can help to realize the linear heavy quark potential. For the hadron spectra, based on the hard-wall AdS/QCD model \cite{EKSS2005} and the soft-wall AdS/QCD or KKSS model \cite{Karch:2006pv}, much effort has been made \cite{Colangelo:2008us, Gherghetta-Kapusta-Kelley,YLWu,Li:2012ay} to realize the spontaneously chiral symmetry breaking and linear confinement properties in hadron spectra. A dynamical holographic QCD (DhQCD) model has been developed in the systematic graviton-dilaton-scalar framework \cite{Li:2012ay}, with the dilaton background field $\Phi(z)$ and the scalar field $X(z)$ describing nonperturbative gluodynamics and chiral dynamics, respectively.
The metric structure at IR in the DhQCD model can be automatically deformed by the nonperturbative gluon condensation and chiral condensation in the vacuum, and the model is quite successful in describing hadron spectra \cite{Li:2013oda}, QCD equation of state \cite{Li:2011hp}, QCD phase transitions and transport properties \cite{Li:2014hja,Li:2014dsa}.

Here in this work, we only focus on the pure gluon system, which can be described by the quenched dynamical holographic QCD model in the graviton-dilaton framework. The action in the string frame takes the form of:
\begin{equation}\label{action-graviton-dilaton}
 S_G=\frac{1}{16\pi G_5}\int
 d^5x\sqrt{g_s}e^{-2\Phi}\left(R_s+4\partial_M\Phi\partial^M\Phi-V^s_G(\Phi)\right).
\end{equation}
Here $G_5$ is the 5D Newton constant, and $g_s$, $\Phi(z)$ and $V_G^s$ are the 5D metric, the dilaton field and dilaton potential in the string frame, respectively.  The dilaton field takes the quadratic form of  $\Phi(z)=\mu_G^2z^2$ to produce linear confinement at IR \cite{Karch:2006pv}. This quenched DhQCD model can describe well not only the scalar glueball spectra \cite{Li:2013oda}, but also all two-gluon and three-gluon glueballs including vectors and tensors \cite{Chen:2015zhh}.

\subsection{First order Hawking-Page phase transition of confinement-deconfinement}

The gauge theory at finite temperature has a holographic counterpart in the thermodynamics of black-holes on the gravity side.  Adding the black-hole background to the quenched dynamical holographic QCD model constructed from vacuum properties, the metric in the string frame takes the form of
\begin{equation} \label{metric-stringframe}
ds_S^2=e^{2A_s}\left(-f(z)dt^2+\frac{dz^2}{f(z)}+dx^{i}dx^{i}\right).
\end{equation}
However, the thermodynamical properties including phase transitions and equation of state are convenient to be derived in the Einstein frame, which is described by
\begin{eqnarray} \label{metric-Einsteinframe}
ds_E^2= e^{2A_s-\frac{4\Phi}{3}}\left(-f(z)dt^2+\frac{dz^2}{f(z)}+dx^{i}dx^{i}\right).
\label{Einstein-metric}
\end{eqnarray}
Under the transformation between the string fame and the Einstein frame as following:
\begin{equation}
g^E_{mn}=g^s_{mn}e^{-2\Phi/3}, ~~ V^E_G=e^{4\Phi/3}V_{G}^s,
\end{equation}
the action at string frame Eq.(\ref{action-graviton-dilaton}) becomes
\begin{eqnarray}\label{graviton-dilaton-E}
S_G^E=\frac{1}{16\pi G_5}\int d^5x\sqrt{g_E}\left(R_E-\frac{4}{3}\partial_m\Phi\partial^m\Phi-V_G^E(\Phi)\right)
\end{eqnarray}
in the Einstein frame.

\begin{figure} [!h]
\centering
\includegraphics[width=0.65\textwidth]{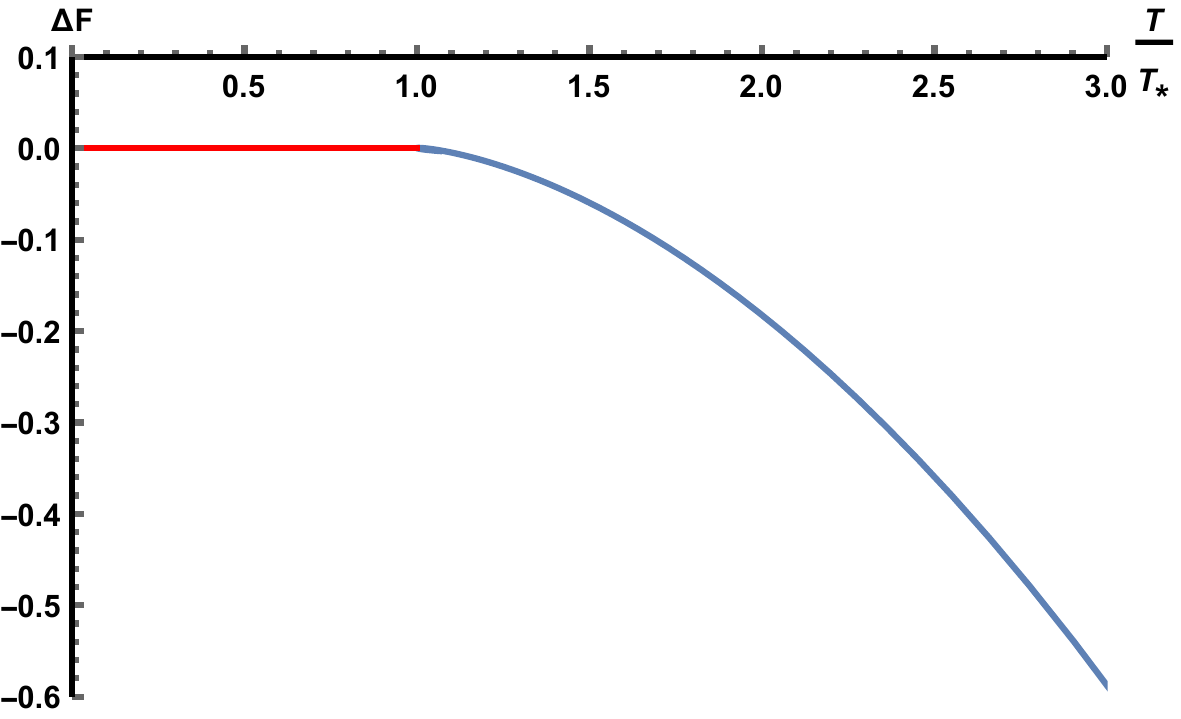}
\caption{The free energy difference of the quenched dynamical holographic QCD model as a function of $T/T_*$. The parameters $\mu_G=0.75 {\rm GeV}$ and $G_5=1.25$ are used and the critical temperature is $T_*=255 {\rm MeV}$.}
\label{fig:energydiffqcd}
\end{figure}

One can derive the following equations of motion (EOMs):
\begin{eqnarray}
 & & -A_s^{''}+A_s^{'2}+\frac{2}{3}\Phi^{''}-\frac{4}{3}A_s^{'}\Phi^{'}=0, \nonumber \\
 & & f''(z)+\left(3 A_s'(z) -2 \Phi '(z)\right)f'(z)=0, \nonumber \\
 & & \frac{8}{3} \partial_z \left(e^{3A_s(z)-2\Phi} f(z) \partial_z \Phi\right)- e^{5A_s(z)-\frac{10}{3}\Phi}\partial_\Phi V_G^E=0,
\end{eqnarray}
and the solution of the black-hole background takes the form of
\begin{eqnarray} \label{solu-f}
f(z)= 1- f_{c}^h \int_0^{z} e^{-3A_s(z^{\prime})+2\Phi(z^{\prime})} dz^{\prime},
\end{eqnarray}
with
\begin{eqnarray}\label{fc}
f_{c}^h= \frac{1}{\int_0^{z_h} e^{-3A_s(z^{\prime})+2\Phi(z^{\prime})} dz^{\prime} }.
\end{eqnarray}
We have $f(z_h)=0$ at the horizon $z=z_h$. The periodicity of the Euclidean time
\begin{equation}
\tau \rightarrow \tau+\frac{4\pi}{|f'(z_h)|}
\end{equation}
determines the temperature of the solution as
\begin{equation}
T=\frac{|f'(z_h)|}{4\pi},
\end{equation}
then one can easily find the relation between the temperature and position of the black hole horizon,
\begin{equation} \label{temp}
T =\frac{e^{-3A_s(z_h)+2\Phi(z_h)}}{4\pi \int_0^{z_h} e^{-3A_s(z^{\prime})+2\Phi(z^{\prime})} dz^{\prime} }.
\end{equation}
With the parameters $\mu_G=0.75 {\rm GeV}$ and $G_5=1.25$ used in \cite{Li:2014hja}, we can get the critical temperature
$T_*=255 {\rm MeV}$ for the first order Hawking-Page confinement-deconfinement phase transition.
The first order phase transition behavior can be also read from the free energy difference $\Delta F = F^{BH}-F_{TG}$ as a function of the temperature shown in Fig.\ref{fig:energydiffqcd}.  Above the critical temperature $T>T_*$, by using the saddle point approximation, the free energy density for the black-hole  has
the form of
\begin{eqnarray}
F^{BH}\simeq TS^{BH}=\frac{T}{\kappa_5^2}\int _0^{T^{-1}} dt\int _{\epsilon}^{z_h}dz~e^{5A(z)-2\Phi(z)}\frac{2}{3}V(\phi)^{BH},
\end{eqnarray}
and below the critical temperature $T<T_*$, the free energy density for the thermal gas takes the form of
\begin{eqnarray}
F_{TG}\simeq TS_{TG}=\frac{T}{\kappa_5^2}\int _0^{\beta '} dt\int _{\epsilon}^{z_{IR}}dz~e^{5A(z)-2\Phi(z)}\frac{2}{3}V(\phi)_{TG}.
\end{eqnarray}
As we can see that the free energy difference $\Delta F$ keeps zero below the critical temperature $T<T_*$, and in the region above the critical temperature $T>T_*$, the free energy difference decreases monotonically with the temperature, which is a typical behavior for first order phase transition.

\section{The first order EW phase transition}
\label{sec:setup-tc}

\subsection{The soft-wall holographic technicolor model}

In the SM, the electroweak phase transition is a crossover \cite{Kajantie:1996mn,Gurtler:1997hr,Csikor:1998eu}. However, when the technicolor models or more general strongly coupled scenarios are considered, (strongly) first-order phase transition at the electroweak or TeV scale is possible. We apply the gauge/gravity duality to model the strong dynamics of electroweak on five dimensional (5D) anti-de Sitter spacetime (AdS$_5$). Similar to the bottom-up holographic model of QCD \cite{Erlich:2005qh,DaRold:2005vr}, we construct a 5D holographic technicolor model or a phenomenological ``$Dp-Dq$" model with the total action of
\begin{eqnarray}
S_{TC}=S_G+S_5.
\end{eqnarray}
$S_G$ is the background ``$Dp$" brane action
\begin{eqnarray}
S_G=\frac{1}{2\kappa_5^2}\int d^5x \sqrt{g}~e^{-w(z)}~\Big(R+\frac{12}{L^2}\Big),
\label{action-b}
\end{eqnarray}
and the background is described by the anti-de Sitter spacetime (AdS$_5$) metric
\begin{equation}
ds^2= g_{MN} dx^M dx^N= \left(L/z \right)^2\big(\eta_{\mu\nu}dx^\mu dx^\nu+dz^2\big),
\end{equation}
with $R$ the Ricci scalar, $L$ the curvature radius of AdS$_5$ and the negative cosmological constant $-12/L^2$.
The probe flavor ``$Dq$" brane with $SU(N_{{\rm TF}})_L \times SU(N_{{\rm TF}})_R$ gauge symmetry ($N_{\rm TF}$ the number of techni-flavors) living on the (AdS$_5$) background is described by the action $S_5$, which takes the form of
\begin{eqnarray}
\label{action-5}
S_5=-\int d^4x\int_{\epsilon}^{z_m}dz \sqrt{-g}~e^{-w(z)} \mathrm{Tr}\Big(|(D^MU)^\dagger(D_MU)|+m_{5}^2|U^\dagger U|\nonumber\\
+\frac{1}{4g_5^2}(L^{MN}L_{MN}+R^{MN}R_{MN})\Big).
\end{eqnarray}
Where the covariant derivative is defined as $D_MU=\partial_M U-iL_MU+iUR_M$, $L(R)_{MN}$ is defined as $L(R)_{MN} = \partial_M L(R)_N - \partial_N L(R)_M-i[L(R)_M, L(R)_N]$, and the gauge coupling $g_5$ is fixed by the UV asymptotic forms of the vector current two-point function as $g_5^2=12\pi^2L/N_{\rm TC}$ \cite{Erlich:2005qh,Haba:2010hu,Matsuzaki:2012xx} with $N_{\rm TC}$ the number of colors. The fifth coordinate is restricted from the infrared (IR) cut-off $z_m$ to the ultraviolet (UV) cut-off $\epsilon$, and we always imply the limit of \(\epsilon\to 0\) for simplicity. In Eqs.(\ref{action-b}) and (\ref{action-5}), we have introduced a soft-wall $w(z)$ to break the scale symmetry breaking. It should be noticed that here $w(z)$ is just a soft cut-off not a dynamical field, therefore, there is no corresponding potential in Eq.(\ref{action-b}) thus $w(z)$ is not needed to be solved from the Einstein equation.

The bulk scalar field $U$ is dual to the boundary operator $\langle \bar{Q}_{{\rm TC}}Q_{{\rm TC}} \rangle$  the chiral flavor symmetry in the boundary theory corresponds to left- and right-gauge fields in the bulk. The 5D mass parameter is related to $m_5^2=-(3-\gamma_m)(1+\gamma_m)/L^2$, and $\gamma_m \simeq 1$ is the same setup as \cite{Haba:2010hu,Matsuzaki:2012xx}. The bulk scalar fields $U$ can be decomposed as
\begin{eqnarray}
U=\frac{1}{\sqrt{2N_{\rm TF}}}\Big(v(z)+\sigma(x,z)\Big)e^{2i\Pi^a(x,z)t^a}.
\end{eqnarray}
Expanding (\ref{action-5}), we find the coupled equations of motion for the vacuum expectation values $v(z)$,
\begin{eqnarray}
\label{eqs-v}
-\frac{1}{e^{-w(z)}}\partial_z(\frac{L^3}{z^3}e^{-w(z)}\partial_zv(z))+\frac{L^5}{z^5}m_{5}^2v(z)=0. \\
\end{eqnarray}
We choose the UV boundary condition for $v(z)$ as
\begin{eqnarray}
v(z)|_{z\to \epsilon}=M z^2,
\end{eqnarray}
where $M$ stands for the current mass of techni-quarks.

Next we expand all the fields in terms of Kaluza-Klein(KK) modes. The Higgs boson in standard model corresponds to the lowest KK mode of bulk scalar $U$ in technicolor scenario named techni-dilaton. Expanding the scalar field $\sigma(x,z)=\sum_n f_{\sigma}^{(n)}(x)\sigma^{(n)}(z)$, the equations of motion for the $\sigma_n(z)$ is,
\begin{eqnarray}
-\frac{z^3}{L^3e^{-w(z)}}\partial_z(\frac{L^3e^{-w(z)}}{z^3}\partial_z\sigma_n)+\frac{L^2}{z^2}m_{5}^2\sigma_n=m_n^2\sigma_n.
\end{eqnarray}
Similar to the \cite{Erlich:2005qh}, we choose their boundary conditions as
\begin{eqnarray}
\partial_z \sigma_n(z_m)=0,~~~\sigma_n(\epsilon)=0.
\end{eqnarray}
We introduce the vector and axial-vector fields as $V_M = (L_M+R_M)/2 $ and $A_M=(L_M-R_M)/2$. In $V_z = A_z =0$ gauge, the equations of motion for the transverse part of the gauge field are
\begin{eqnarray}
-\frac{z}{e^{-w(z)}L}\partial_z(\frac{e^{-w(z)}L}{z}\partial_z V_n^a)=m_n^2V_n^a,\\
-\frac{z}{e^{-w(z)}L}\partial_z(\frac{e^{-w(z)}L}{z}\partial_z A_n^a)+\frac{v^2 g^2L^2}{z^2}A_n^a=m_n^2A_n^a.
\end{eqnarray}
The boundary conditions are chosen as:
\begin{eqnarray}
\partial_z V_n^a(z_m)=\partial_z A_n^a(z_m)=0,\\
V_n^a(\epsilon)=0,~~~A_n^a(\epsilon)=0.
\end{eqnarray}
It is similar to the holographic QCD case that we can get a set of vector and axial-vector mesons which the lowest eigenvalue is identified as the techni-$\rho$ and techni-$a_1$ meson, respectively. The pseudo-scalar field is coupled with longitudinal part of axial-vector field.
The resulting equations of motion are($A_{\mu}=A_{\mu \perp}+\partial_{\mu}\varphi$)
\begin{eqnarray}
\partial_z(\frac{e^{-w(z)}L}{z}\partial_z\varphi_n^a)+\frac{g_5^2v^2e^{-w(z)}L^3}{z^3}(\Pi_n^a-\varphi_n^a)=0,\\
-m_n^2\partial_z\varphi_n^a+\frac{g_5^2v^2e^{-w(z)}L^2}{z^2}\partial_z\Pi_n^a=0.
\end{eqnarray}
We choose their boundary conditions as:
\begin{eqnarray}
\partial_z\varphi_n^{a}(z_m)=\varphi_n^a(\epsilon)=\Pi_n^a(\epsilon)=0.
\end{eqnarray}
The techni-pion decay constant and $S$ parameter is given as \cite{Haba:2010hu,Matsuzaki:2012xx}
\begin{eqnarray}
\label{pion-dc}
f_{\Pi}^2=-\frac{1}{g_5^2}\frac{L}{z}e^{-w(z)}\partial_zA(0,z)|_{z=\epsilon},\\
S=N_D 4\pi\frac{L}{g_5^2}\int_{\epsilon}^{z_m}dz\frac{1}{z}e^{-w(z)}(1-A(0,z)).
\label{ptp-s}
\end{eqnarray}
Where $N_D$ is the number of generation.

Similar to the soft-wall holographic QCD model in Ref.\cite{Karch:2006pv}, we assume the soft-wall has the quadratic form $w(z)=-c z^2$. There are six free parameters $N_{\rm TF}$, $N_{\rm TC}$, $N_D$, $z_m$, $M$ and $c$ in our model. For simplicity, we fix the $N_{\rm TF}=2$ and $N_D=1$ as the minimal technicolor model, then only four free parameters are left. In order to produce realistic mass spectra, we choose the Higgs boson mass, techni-pion decay constant and the $S$ parameter as input to trade off with the parameters ($z_m$, $M$, $c$). Then once the value of $N_{\rm TC}$ is fixed, all spectra can be found. According to the PDG \cite{Patrignani:2016xqp}, the electroweak precision tests put a constraint to the S parameter as $-0.01 \leq S \leq 0.15$ (at 90\% CL) assuming another oblique parameter $U = 0$. The technicolor scenario requires the techni-pion decay constant $f_{\Pi}=246{\rm GeV}$.
As shown in Table \ref{tab:spectum}, for Model I, by fitting to three experimental data $S=0.15$, $m_{\rm higgs}=125 {\rm GeV}$ and $f_{\Pi}=246{\rm GeV}$, three free parameters ($z_m$, $M$, $c$) can be found with a fixed value of $N_{\rm TC}=3,4,5$, respectively.  On this basis, we obtained the techni-mesons spectra for $N_{\rm TC}=3$, 4 and 5, respectively. We find that in by using the parameters set of Model I all the masses of techni-mesons are heavier than 2TeV.

\begin{table}[!h]
\begin{tabular}{|c|c|c|c|c|c|c|c|}
\hline
\multicolumn{2}{|c|}{\multirow{2}*{~}} & \multicolumn{3}{|c|}{Model I} & \multicolumn{3}{|c|}{Model II} \\
\cline{3-8}
\multicolumn{2}{|c|}{~} & $N_{\rm TC}=3$ & $N_{\rm TC}=4$ & $N_{\rm TC}=5$ & $N_{\rm TC}=3$ & $N_{\rm TC}=4$ & $N_{\rm TC}=5$ \\
\cline{1-8}
\multirow{3}{*}{\begin{sideways}{\textbf{Input~~}}\end{sideways}} & $z_m^{-1}$(TeV) & 1.072 & 1.079 & 1.083 & 0.299 & 0.299 & 0.299 \\
\cline{2-8}
~ & $c$ & 1.395 & 1.414 & 1.423 & 0.099 & 0.099 & 0.099 \\
\cline{2-8}
~ & $M$ & 0.553 & 0.541 & 0.534 & 0.380 & 0.329 & 0.295 \\
\cline{1-8}
\multirow{8}{*}{\begin{sideways}{\textbf{Output~~~~~~~}}\end{sideways}} & $S$ & \multicolumn{3}{|c|}{0.15(fixed)} & 0.721 & 0.888 & 1.034 \\
\cline{2-8}
~ & $T_*$(GeV) & 356 & 358 & 360 & \multicolumn{3}{|c|}{100(fixed)} \\
\cline{2-8}
~ & $m_{{\rm higgs}}^{(1)}$ & 4.368 & 4.397 & 4.413 & 1.217 & 1.217 & 1.217 \\
~ & $m_{{\rm techni}-\rho}^{(0)}$ & 2.054 & 2.068 & 2.075 & 0.586 & 0.586 & 0.586 \\
~ & $m_{{\rm techni}-\rho}^{(1)}$ & 5.722 & 5.759 & 5.780 & 1.603 & 1.603 & 1.603 \\
~ & $m_{{\rm techni}-a_1}^{(0)}$ & 2.342 & 2.274 & 2.236 & 2.853 & 2.292 & 1.882 \\
~ & $m_{{\rm techni}-a_1}^{(1)}$ & 5.814 & 5.824 & 5.830 & 3.090 & 2.515 & 2.242 \\
~ & $m_{\Pi}^{(1)}$ & 5.531 & 5.521 & 5.515 & 3.135 & 2.686 & 2.385 \\
\hline
\end{tabular}
\caption{Model I fits three experimental data $S=0.15$, $m_{\rm higgs}^{(0)}=125 {\rm GeV}$ and $f_{\Pi}=246{\rm GeV}$. Model II fits two experimental data $m_{\rm higgs}^{(0)}=125 {\rm GeV}$, $f_{\Pi}=246{\rm GeV}$ and assumes a phase transition temperature of 100 GeV. Note that the unit of particle mass is TeV.}
\label{tab:spectum}
\end{table}

\subsection{The first order Hawking-Page phase transition for EW}

We have assumed that the flavor brane is a probe, therefore the thermodynamical properties of the system is dominated by the background action $S_G$ in Eq.(\ref{action-b}). The thermal AdS solution in Poincar$\acute{\mathrm{e}}$ patch is
\begin{eqnarray}
ds^2=\frac{L^2}{z^2}\Big(dt^2+d\vec{x}^2+dz^2\Big),~~~\epsilon\leq z\leq z_m.
\end{eqnarray}
The temperature of thermal AdS is $T=1/\beta '$, where $ \beta ' $ is the period of the Euclidean time.
The AdS-BH solution is
\begin{eqnarray}
ds^2=\frac{L^2}{z^2}\Big(f(z)dt^2+d\vec{x}^2+\frac{dz^2}{f(z)}\Big),~~~\epsilon\leq z\leq {\rm min}(z_m, z_h),
\end{eqnarray}
with $f(z)=1-z^4/z_h^4$. The Hawking temperature of the black hole is $T=1/(\pi z_h)$.

By choosing the soft-wall in the form of $w(z)=-c z^2$, we can compute the free energy for thermal AdS and AdS black hole
\begin{eqnarray}
F^{AdS}&\simeq&\frac{4L^3T}{\kappa_5^2}\int _0^{\beta '} dt\int _{\epsilon}^{z_m}dz~z^{-5}~e^{c z^2} \nonumber \\
&=& \frac{4L^3T}{\kappa_5^2}\left[-\frac{e^{c z_m^2}}{z_m^4}-\frac{c e^{c z_m^2}}{z_m^2}
+\frac{e^{c z_m^2}}{\epsilon^4}+\frac{c e^{c z_m^2}}{\epsilon^2}+c^2({\rm Ei}(c z_m^2)-{\rm Ei}(c \epsilon^2))\right],\\
F^{BH}&\simeq&\frac{4L^3T}{\kappa_5^2}\int _0^{\pi z_h} dt\int _{\epsilon}^{{\rm min}(z_m, z_h)}dz~z^{-5}~e^{c z^2} \nonumber \\
&=& \frac{4L^3T}{\kappa_5^2}[-\frac{e^{c ({\rm min}(z_m, z_h))^2}}{({\rm min}(z_m, z_h))^4}-\frac{\mu e^{c ({\rm min}(z_m, z_h))^2}}{({\rm min}(z_m, z_h))^2}\nonumber\\
&+&\frac{e^{c ({\rm min}(z_m, z_h))^2}}{\epsilon^4}+\frac{c e^{\mu ({\rm min}(z_m, z_h))^2}}{\epsilon^2}+c^2({\rm Ei}(c ({\rm min}(z_m, z_h))^2)-{\rm Ei}(c \epsilon^2))],
\end{eqnarray}
respectively.  The free energy difference is
\begin{eqnarray}
\Delta F=\left \{
\begin{array}{lc}
\frac{L^3}{\kappa_5^2}
\frac{1}{2 z_h^4} & z_m < z_h\\
\frac{L^3}{\kappa_5^2}
\left[\frac{1-2e^{c z_h^2}(1+c z_h^2)}{2z_h^4}+\frac{e^{c z_m^2}(1+c z_m^2)}{z_m^4}+c^2({\rm Ei}(c z_h^2)-{\rm Ei}(c z_m^2))\right] & z_m > z_h,\\
\end{array}
\right. \
\end{eqnarray}
where Ei(x)$\equiv\int_{-x}^\infty dt~e^{-t}/t$.
\begin{figure} [!h]
\centering
\includegraphics[width=0.45\textwidth]{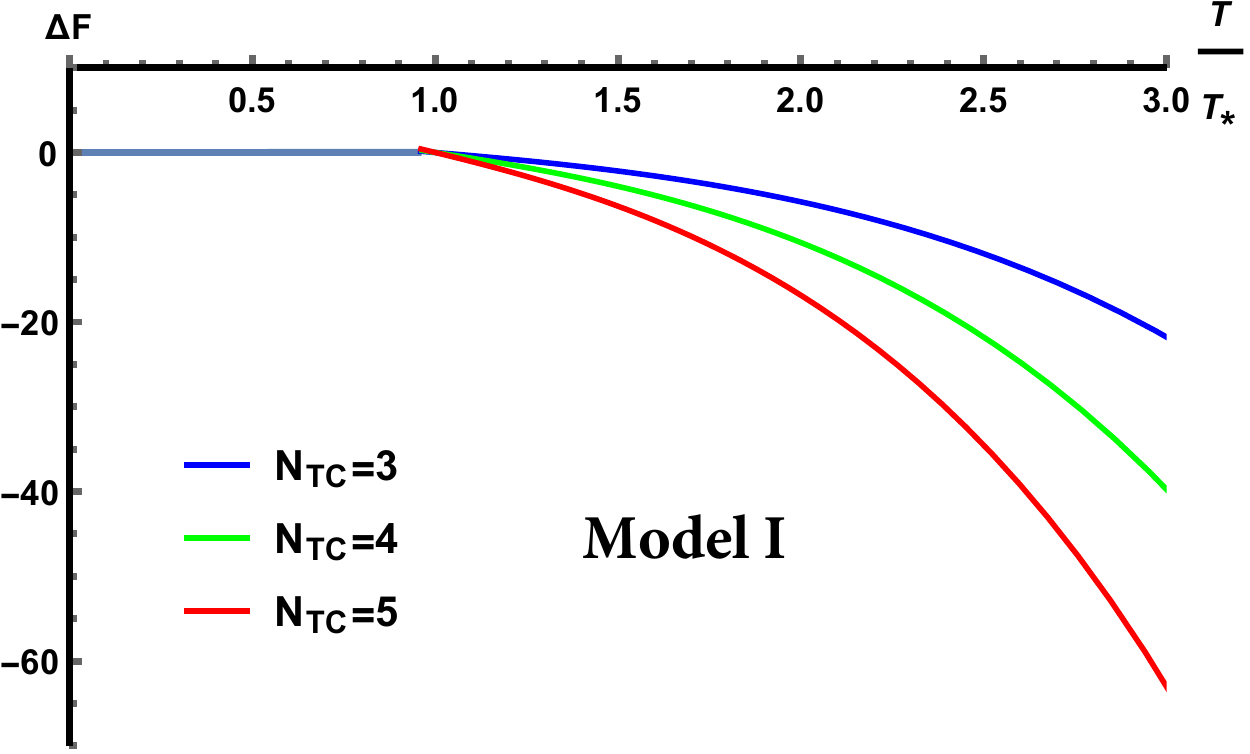}
\includegraphics[width=0.45\textwidth]{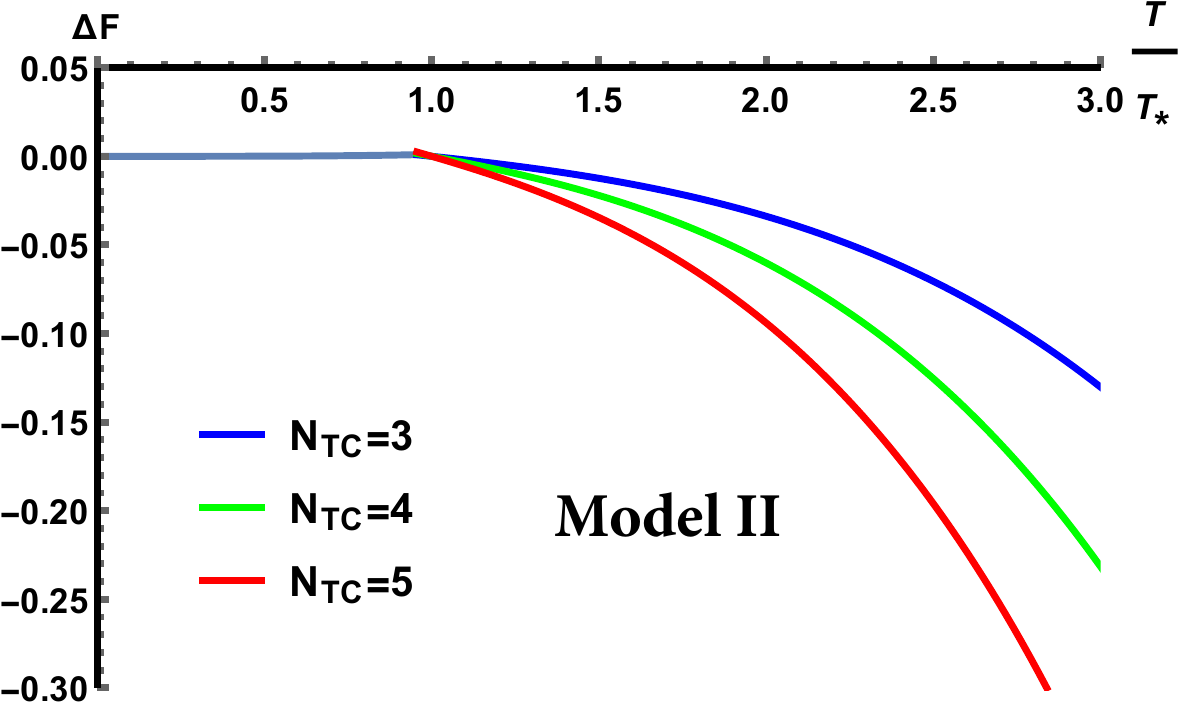}
\caption{The free energy difference for Model-I and Model-II with $N_{\rm TC}=3$, $N_{\rm TC}=4$ and $N_{\rm TC}=5$, respectively.}
\label{fig:frenergyew}
\end{figure}

The 1st-order Hawking-Page phase transition occurs when free energy difference changes the sign as shown in Fig.\ref{fig:frenergyew}. For Model I in Table \ref{tab:spectum}, the EW phase transition happens at the critical temperature $T_*=356, 358, 360 {\rm GeV}$ for $N_{TC}=3,4,5$, respectively. Considering the typical critical temperature for EW phase transition is around $80-150 {\rm GeV}$, by fixing the critical temperature at $T_*=100 {\rm GeV}$, and fitting $m_{\rm higgs}^{(0)}=125 {\rm GeV}$, $f_{\Pi}=246{\rm GeV}$, we can have a rather large $S$ parameter $0.721, 0.888, 1.034 $ for $N_{TC}=3,4,5$, respectively, which can be read from Model II of Table  \ref{tab:spectum}. On this basis, we obtained the techni-mesons spectra for $N_{\rm TC}=3$, 4 and 5, respectively. We find that comparing with Model I, the masses of techni-mesons in Model II are much lighter.

\section{Gravitational wave}
\label{sec:gw}

When the first-order phase transition takes place, the bubbles can nucleate in the supercooling plasma and the latent heat of free energy between symmetric and broken phases can engine the expansion of the bubble walls. Gravitational waves can be generated by the bubbles collisions \cite{Huber:2008hg,Jinno:2016vai}, by the stirred acoustic waves \cite{Hindmarsh:2015qta,Hindmarsh:2017gnf} and by the Magnetohydrodynamic(MHD) turbulence in the plasma  \cite{Caprini:2009yp,Binetruy:2012ze} (see \cite{Caprini:2015zlo,Weir:2017wfa} for review).

The gravitational waves power spectra are consisted of these three components, which can be put as
\begin{eqnarray}
h^2\Omega (f)=h^2\Omega_\text{env}(f)+h^2\Omega _{sw}(f)+h^2\Omega _{turb}(f).
\end{eqnarray}
The gravitational wave power spectrum from bubble collisions in the envelope approximation can be given as $h^2\Omega_\text{env}(f)$, which takes the following form \cite{Huber:2008hg} via numerical calculation
\begin{eqnarray}
h^2 \Omega_\text{env}(f) = 1.67 \times 10^{-5} \, \Delta\left(\frac{H_*}{\beta}\right)^2 \left( \frac{\kappa_\phi \alpha}{1+\alpha} \right)^2 \left(\frac{100}{g_*} \right)^{\frac{1}{3}} S_\text{env}(f) \,,
\end{eqnarray}
where the spectral form $S_\text{env}(f)$ have the power law which is given as
\begin{eqnarray}
S_\text{env}(f) = \frac{3.8(f/f_\text{env})^{2.8}}{1+2.8(f/f_\text{env})^{3.8}}\,.
\end{eqnarray}
Here $\alpha$ is the ratio of vacuum energy density over radiation energy density, $g_*$ is the number of active degrees of freedom during the phase transition,  $v_\mathrm{w}$ is the velocity of bubble wall expansion,
$\kappa_\phi$ is the coefficient which measure the efficiency of converting vacuum energy into scalar field gradient energy
and $H_*/\beta$ is the nucleation rate relative to the Hubble rate during the phase transition which measures the time duration of phase transition.

The peak frequency of the GW produced by bubble collisions can be estimated as
\begin{eqnarray}
f_\text{env} = 16.5  \, \mu\mathrm{Hz} \,\left(\frac{f_*}{\beta} \right) \left( \frac{\beta}{H_*} \right)\left( \frac{T_*}{100 \, \mathrm{GeV}} \right) \left( \frac{g_*}{100}\right)^{\frac{1}{6}}\,,
\end{eqnarray}
where $T_*$ denotes the temperature during the phase transition.
The dependence of the amplitude and peak frequency on the velocity of bubble expansion $v_\mathrm{w}$ is provided as
\begin{eqnarray}
\Delta =\frac{0.11v_\mathrm{w}^3}{0.42+v_\mathrm{w}^2}; \quad\frac{f_*}{\beta} = \frac{0.62}{1.8-0.1v_\mathrm{w}+v_\mathrm{w}^2}.
\end{eqnarray}

The power spectrum of GW from acoustic waves can be found in reference \cite{Hindmarsh:2015qta}, which can take the following form
\begin{eqnarray}
h^2 \Omega_\text{sw}(f) =2.65 \times 10^{-6} v_\mathrm{w} \left(\frac{H_*}{\beta}\right)   (\frac{\kappa_\mathrm{f}\alpha}{1+\alpha})^2 \left(\frac{100}{g_*} \right)^{\frac{1}{3}} \, S_\text{sw}(f)
\end{eqnarray}
Here $\kappa_\mathrm{f}$ is the ratio of bulk kinetic energy to vacuum energy.  And the spectral shape is given
\begin{eqnarray}
S_\text{sw}(f) = \left(\frac{f}{f_\mathrm{sw}}\right)^3  \left(\frac{7}{4 + 3 (f/f_\mathrm{sw})^2 } \right)^{7/2}
\end{eqnarray}
with approximate peak frequency
\begin{eqnarray}
f_\mathrm{sw} = 19\times 10^{-6} \, \mathrm{Hz} \, \frac{1}{v_\mathrm{w}} \left( \frac{\beta}{H_*} \right) \left( \frac{T_*}{100 \, \mathrm{GeV}}\right) \left( \frac{g_*}{100} \right)^\frac{1}{6},
\end{eqnarray}

The gravitational wave power spectrum from Kolmogorov-type turbulence is \cite{Caprini:2009yp,Binetruy:2012ze}
\begin{eqnarray}
h^2 \Omega_\text{turb} (f) = 3.35 \times 10^{-4} v_\mathrm{w} \left(\frac{H_*}{\beta} \right) \left( \frac{\kappa_\text{turb} \alpha}{1+ \alpha} \right)^{\frac{3}{2}} \left( \frac{100}{g_*}\right)^{\frac{1}{3}}  S_\text{turb}(f).
\end{eqnarray}
Here $\kappa_\text{turb}$ is the efficiency of conversion of latent heat into turbulent flows. The spectral shape of the turbulent contribution is
\begin{eqnarray}
S_\text{turb}(f) = \frac{(f/f_\text{turb})^3}{[1+(f/f_\text{turb})]^{\frac{11}{3}} ( 1+ 8 \pi f/h_*) }
\end{eqnarray}
where $h_*$ is the Hubble rate at $T_*$,
\begin{eqnarray}
h_* = 16.5 \, \mu\mathrm{Hz} \left( \frac{T_*}{100 \, \mathrm{GeV}}\right)  \left(\frac{g_*}{100} \right)^{\frac{1}{6}}.
\end{eqnarray}
The peak frequency $f_\text{turb}$ is
\begin{eqnarray}
f_\text{turb} = 27 \, \mu\mathrm{Hz} \, \frac{1}{v_\mathrm{w}} \left( \frac{\beta}{H_*} \right)\left(\frac{T_*}{100 \, \mathrm{GeV}}  \right) \left(\frac{g_*}{100}\right)^{\frac{1}{6}}.
\end{eqnarray}
It is a remarkable fact that four factors $\kappa_\phi$, $\kappa_\mathrm{f}$, $\kappa_\text{turb}$ and $v_\mathrm{w}$ are model-dependent. If we focus on the case of Jouguet detonations \cite{Steinhardt:1981ct,Kamionkowski:1993fg,Nicolis:2003tg,Espinosa:2010hh}, then
\begin{eqnarray}
\kappa_\phi&=&\frac{1}{1+0.715\alpha}\bigg(0.715\alpha+\frac{4}{27}\sqrt{\frac{3\alpha}{2}}\bigg),\\
\kappa_\mathrm{f}&=&\frac{\sqrt{\alpha}}{0.135+\sqrt{ \alpha+0.98 }},\\
v_\mathrm{w}&=&\frac{\sqrt{1/3}+\sqrt{\alpha^2+2\alpha/3}}{1+\alpha}.
\end{eqnarray}
We can compute the latent heat and $\alpha$ by
\begin{eqnarray}
\varepsilon_*&=&\Big(-\Delta F(T)+T\frac{d \Delta F(T)}{dT}\Big)\Bigg|_{T=T_*}, \nonumber \\
\alpha&=&\frac{\varepsilon_*}{\pi^2g_*T_*^4/30}.
\end{eqnarray}

The Refs.\cite{Hindmarsh:2015qta,Caprini:2015zlo} suggest that only at most 5-10\% of the bulk motion from the bubble walls is converted into vorticity, so we assume $\kappa_\text{turb}=0.05\kappa_\phi$. Using holographic models for QCD and EW phase transitions, we can calculate the transition temperature $T_*$ and phase transition strength $\alpha$, and obtain the gravitational wave power spectrum.

\begin{table}[!h]
\begin{center}
\begin{tabular}{|c|c|c|c|}
\hline
~ & $\alpha$ & $H_*/\beta$ & $g_*$ \\
\cline{1-4}
QCD & 0.611 & 1/10 & 10 \\
EW Model I $N_{\rm TC}=3$ & 6.885 & 1/100 & 100 \\
EW Model I $N_{\rm TC}=4$ & 12.245 & 1/100 & 100 \\
EW Model I $N_{\rm TC}=5$ & 19.136 & 1/100 & 100 \\
EW Model II $N_{\rm TC}=3$ & 6.037 & 1/100 & 100 \\
EW Model II $N_{\rm TC}=4$ & 10.732 & 1/100 & 100 \\
EW Model II $N_{\rm TC}=5$ & 16.769 & 1/100 & 100 \\
\hline
\end{tabular}
\label{tab:para}
\caption{Parameters $\alpha$, $H_*/ \beta$ and $g_*$ used to calculate the GWs in the holographic QCD model and holographic EW model, respectively.}
\end{center}
\end{table}

\begin{figure} [!h]
\centering
\includegraphics[width=0.65\textwidth]{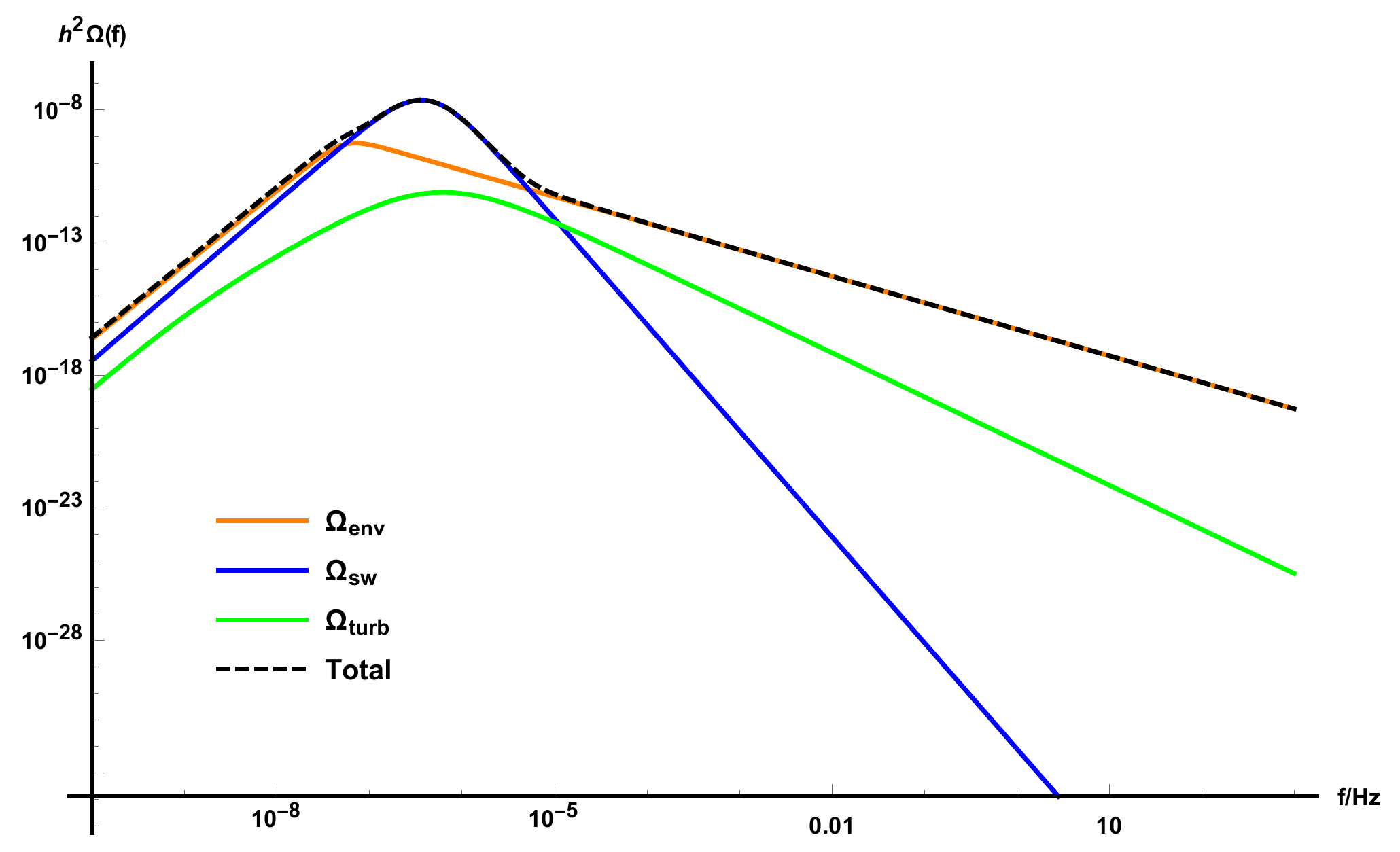}
\caption{The GWs spectrum or the energy density of the GWs produced from the first order QCD phase transition by using the holographic QCD model.}
\label{fig:gwqcd}
\end{figure}

\vskip 0.5 cm
{\bf GWs from QCD phase transition:}
\vskip 0.5 cm

For the dynamical holographic QCD model in Sec.\ref{sec:setup-qcd},  with the parameters $\mu_G=0.75 {\rm GeV}$ and $G_5=1.25$, the critical temperature for the first order phase transition is $T_*=255 {\rm MeV}$, and one can calculate $\alpha=0.611$. Since this holographic QCD model is for pure gluon system and quark dynamics has not been considered, we cannot obtain the specific values of $g_*$ and $H_*/ \beta$. So we impose typical values of $g_*=10$ and $H_*/ \beta=1/10$ for the QCD phase transition \cite{Ahmadvand:2017xrw,Ahmadvand:2017tue} as listed in Table \ref{tab:para}. The gravitational wave power spectra produced from QCD phase transition is shown in Fig.\ref{fig:gwqcd}. It is found that the peak frequency is located around $3\times 10^{-7}$ Hz, and the location of the peak frequency is determined by the sound waves. The bubble collisions contribution to the GWs spectrum is dominant in the frequency region $f<5\times 10^{-8}$ Hz and $f>10^{-5}$ Hz, and the contribution from sound waves to the GWs spectrum is dominated in the region of $f<10^{-5}$ Hz.  The contribution from turbulence to the GWs spectrum is comparably small.

\vskip 0.5 cm
{\bf GWs from EW phase transition:}
\vskip 0.5 cm

\begin{figure} [!h]
\centering
\includegraphics[width=0.32\textwidth]{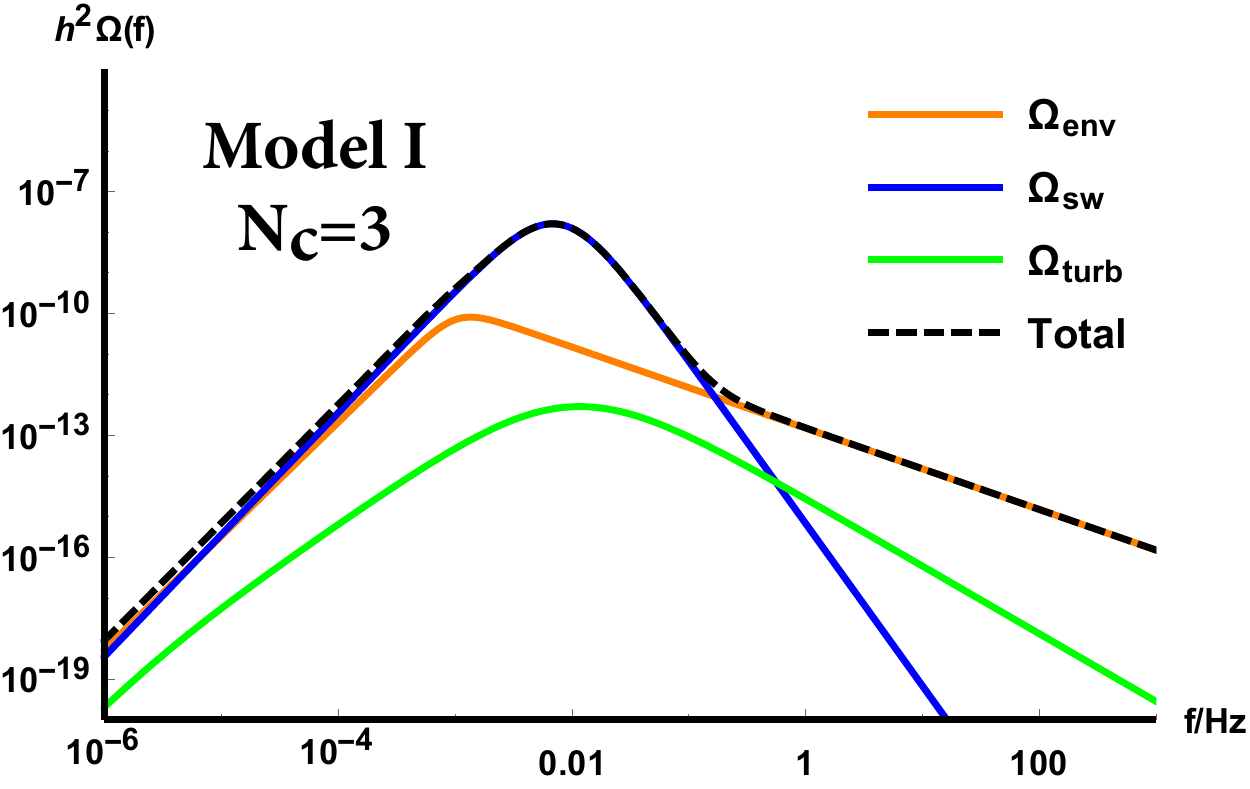}
\includegraphics[width=0.32\textwidth]{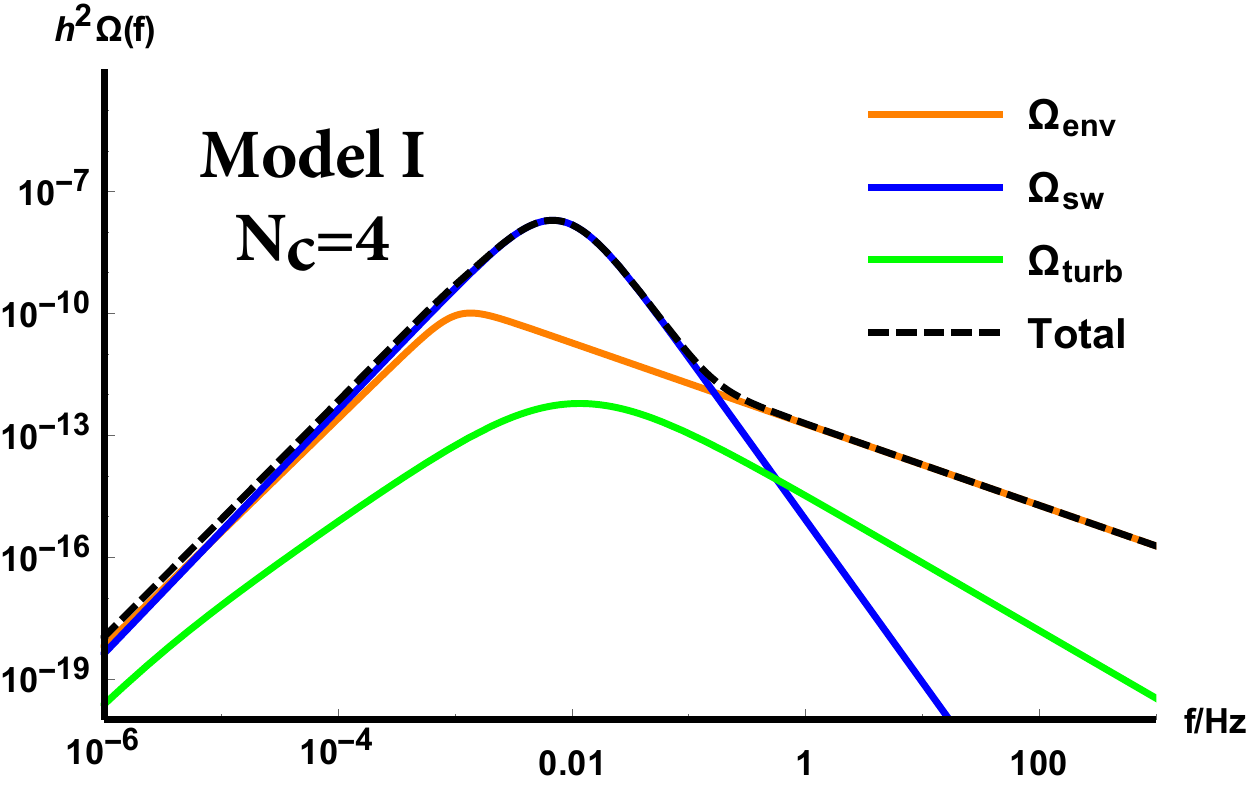}
\includegraphics[width=0.32\textwidth]{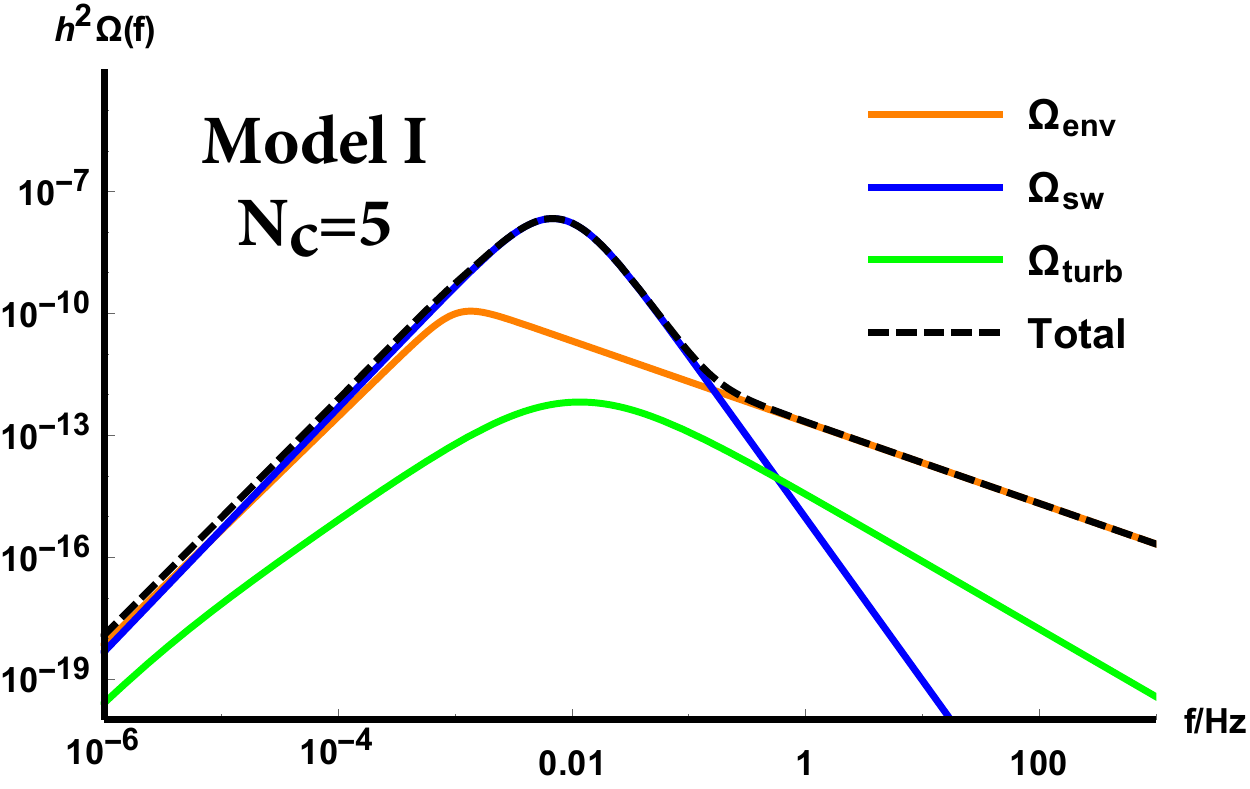}
\caption{The GWs spectrum or the energy density of the GWs as a function of the frequency produced from the first order EW phase transition, with the parameters listed in Model I of the holographic technicolor model for $N_{\rm TC}=3$, $N_{\rm TC}=4$ and $N_{\rm TC}=5$, respectively.}
\label{fig:gwew-I}
\end{figure}

\begin{figure} [!h]
\centering
\includegraphics[width=0.32\textwidth]{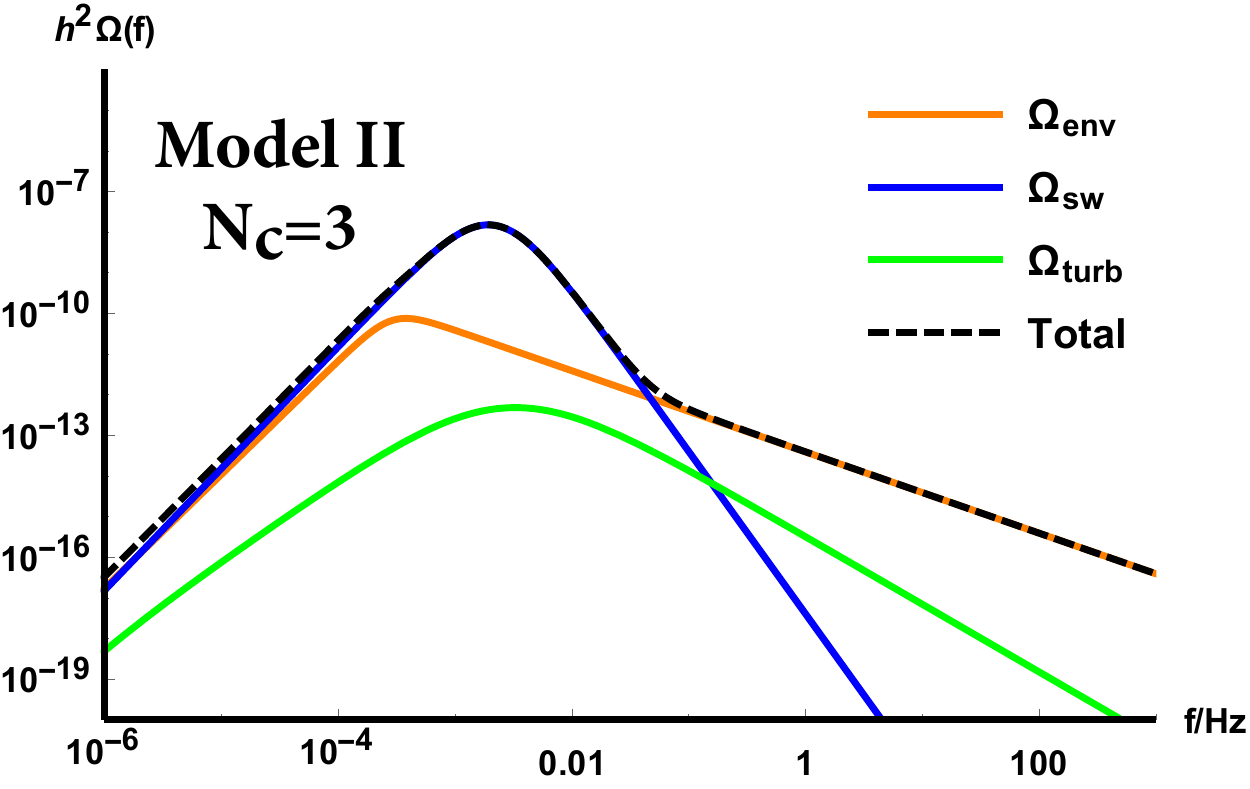}
\includegraphics[width=0.32\textwidth]{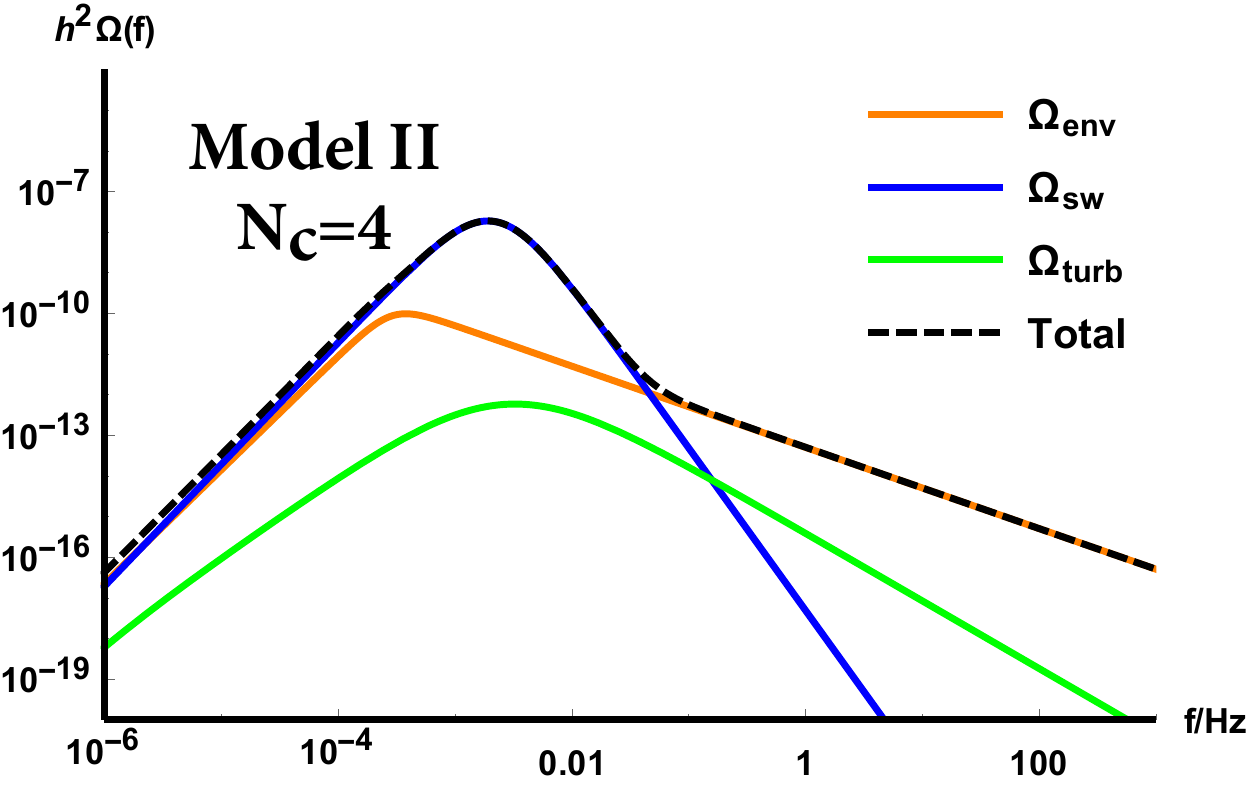}
\includegraphics[width=0.32\textwidth]{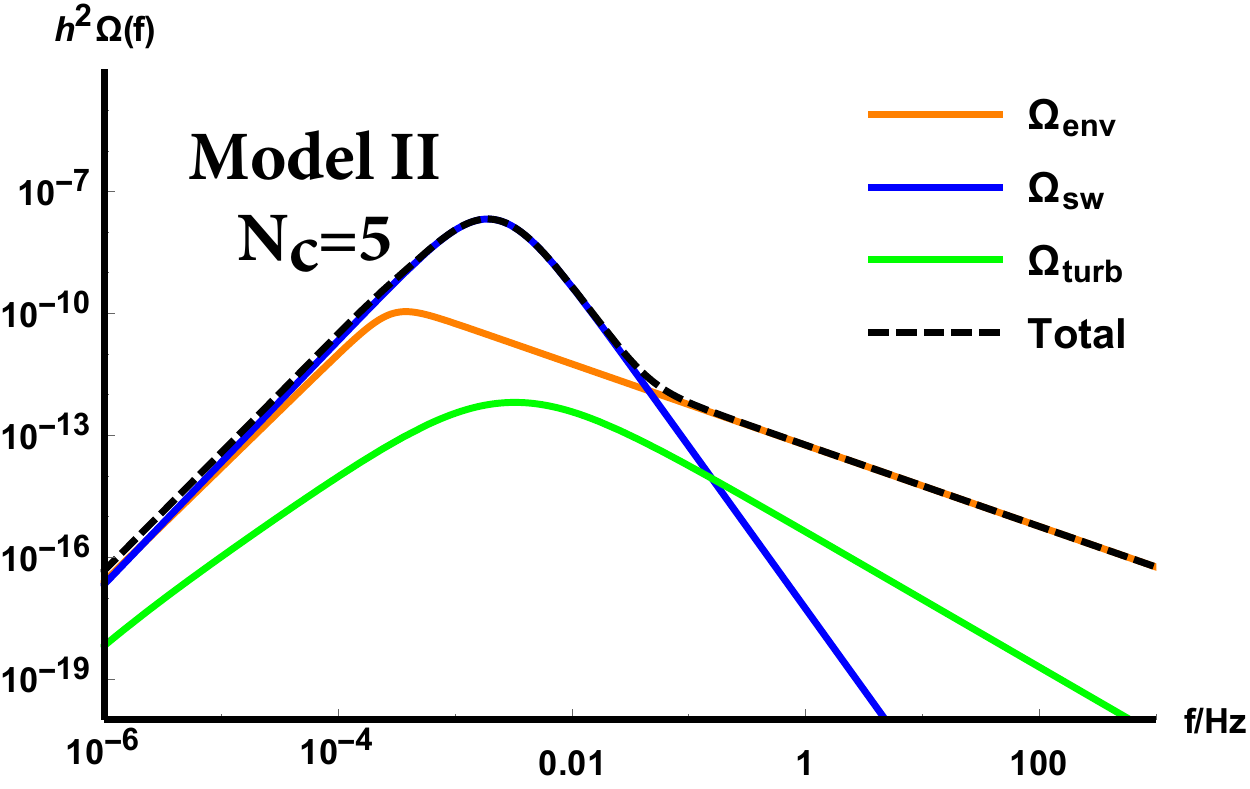}
\caption{The GWs spectrum or the energy density of the GWs as a function of the frequency produced from the first order EW phase transition, with the parameters listed in Model II of the holographic technicolor model for $N_{\rm TC}=3$, $N_{\rm TC}=4$ and $N_{\rm TC}=5$, respectively.}
\label{fig:gwew-II}
\end{figure}

For holographic EW model, the latent heat and $\alpha$ can be derived analytically:
\begin{eqnarray}
\varepsilon_*&=&\Big(-\Delta F(T)+T\frac{d \Delta F(T)}{dT}\Big)\Bigg|_{T=T_*}=\frac{N_{\rm TC}^2\pi^2T_*^4}{2}\Big(2~e^{\frac{c}{\pi^2T_*^2}}-1\Big),\\
\alpha&=&\frac{\varepsilon_*}{\pi^2g_*T_*^4/30}=\frac{3N_{\rm TC}^2}{20}\Big(2~e^{\frac{c}{\pi^2T_*^2}}-1\Big),
\end{eqnarray}
where we use $L^3/\kappa_5^2=N_{\rm TC}^2/(4\pi^2)$. Since the fermion part of 5D action has not been considered, we cannot obtain the specific values of $g_*$ and $H_*/ \beta$, we impose typical values of $g_*=100$ and $H_*/ \beta=1/100$ for EW phase transition. In Fig. \ref{fig:gwew-I} and \ref{fig:gwew-II}, we plot the gravitational wave power spectra for Model I and Model II with parameters listed in Table \ref{tab:spectum} of Sec.\ref{sec:setup-tc}.

For parameters in Model I, the critical temperature for the fist order EW phase transition is at $T_*=356,358,360 {\rm GeV}$ for $N_{TC}=3,4,5$, respectively.
For parameters in Model II, the critical temperature for the fist order EW phase transition is fixed at $T_*=100 {\rm GeV}$ for $N_{TC}=3,4,5$.
From Fig. \ref{fig:gwew-I} and \ref{fig:gwew-II}, we can see that the peak frequency is located at around $f=0.007$ Hz for Model I and around $f=0.002$ Hz for Model II. Similar to the QCD case, the location of the peak frequency is determined by the sound waves. The bubble collisions contribution to the GWs spectrum is dominant in the frequency region $f<10^{-6}$ Hz and $f>0.5$ Hz for Model I and dominant in the frequency region $f<10^{-5}$ Hz and $f>0.2$ Hz for Model II, and the contribution from sound waves to the GWs spectrum is dominated in the region of $f<0.5$ Hz for Model I and in the region of $f<0.2$ Hz for Model II. The contribution from turbulence to the GWs spectrum is comparably small.

\begin{figure} [!h]
\centering
\includegraphics[width=0.8\textwidth]{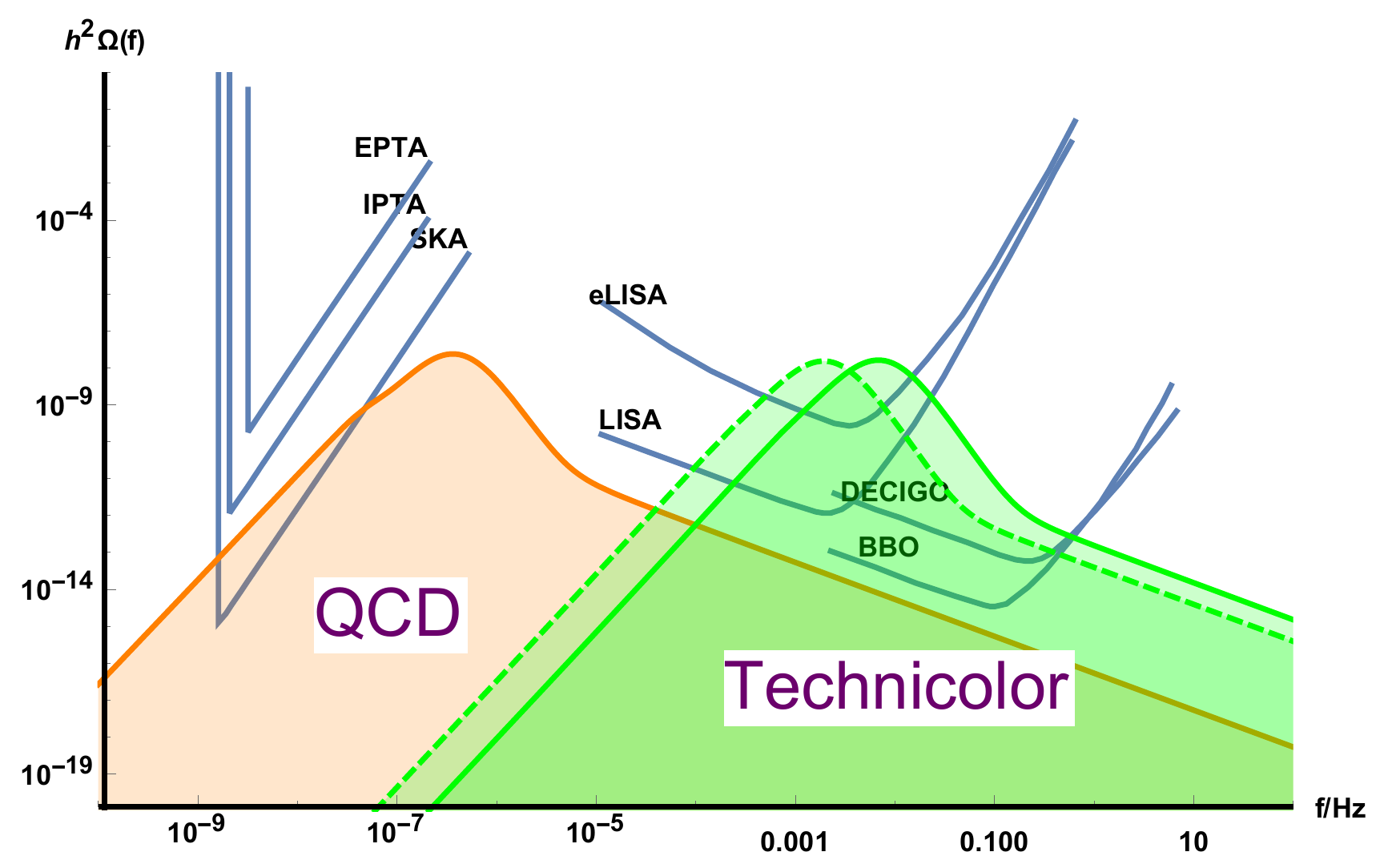}
\caption{The combined GWs spectra from QCD and EW phase transitions and compare with possible GWs detectors. }
\label{fig:gwcp}
\end{figure}

We combine the GWs spectra produced from QCD and EW phase transitions in Fig.\ref{fig:gwcp} and show the frequency and energy density region for different GWs detectors \cite{GW-detector,Kuroda:2015owv}. It can be read that the energy density of GWs produced from first order QCD phase transition can reach $10^{-8}$ around the peak frequency region $3\times 10^{-7}$ Hz, which might be detected by FAST \cite{FAST} and SKA, and the energy density of GWs produced from first order EW phase transition can reach $10^{-8}$ around the peak frequency $0.002 \sim 0.007$ Hz, which can be detected by BBO, DECIGO, LISA and eLISA.

\section{Conclusions and discussions}
\label{sec:sum}

In this work, we have investigated the GWs produced from QCD and EW phase transitions in the framework of holographic QCD model and holographic technicolor model.

The quenched dynamical holographic QCD model describes the pure gluon system, and the first order phase transition for confinement-deconfinement occurs at the critical temperature $255 {\rm MeV}$ with the parameters fixed by glueball spectra. The energy density of the GWs produced from QCD phase transition can reach
$10^{-8}$ around the peak frequency region $3\times 10^{-7}$ Hz, which might be detected by the GWs detector FAST and SKA.

The holographic technicolor model is constructed by mimicking the phenomenological ``Dp-Dq" system, the probe flavor ``$Dq$" brane has $SU(N_{{\rm TF}})_L \times SU(N_{{\rm TF}})_R$ gauge symmetry with $N_{\rm TF}$ the number of techni-flavors living on the (AdS$_5$) background ``$Dp$" brane. The higgs particle and techni-mesons are generated by the probe flavor brane action, and the thermodynamical properties of the system is dominated by the background brane action. For the minimal technicolor model, we have two sets of parameters corresponding to Model I and Model II, respectively. For Model I, by fitting free parameters $z_m$, $M$, $c$ from three experimental data: $S$ parameter, mass of higgs boson and vacuum expectation value of electroweak, the model can have first order EW phase transition at the critical temperature around $350 {\rm GeV}$. For Model II, we fix the critical temperature for EW at $100 {\rm GeV}$.
With the critical temperature ranging from $100 - 360 {\rm GeV}$, the energy density of GWs produced from first order EW phase transition can reach $10^{-8}$ around the peak frequency $0.002 \sim 0.007$ Hz, which can be detected by BBO, DECIGO, LISA and eLISA.

We also observe that for both GWs produced from QCD and EW phase transitions, in the peak frequency region, the dominant contribution comes from the sound waves, and away from the peak frequency region, the contribution from bubble collision is dominant to the GWs.

\vskip 0.5cm
{\bf Acknowledgement}  We thank valuable discussion with F.P. Huang, D.N. Li, Z.B. Li and Y.J.Zhao. M.Huang is supported by the NSFC under Grant No. 11725523, 11735007 and 11261130311(CRC 110 by DFG and NSFC), and Q.S. Yan is supported by the NSFC under the grant NO. 11575005.


\vskip 0.2cm

\begin{thebibliography}{99}

\bibitem{Einstein:1916cc}
 A.~Einstein,
  %``Approximative Integration of the Field Equations of Gravitation,''
  Sitzungsber.\ Preuss.\ Akad.\ Wiss.\ Berlin (Math.\ Phys.\ ) {\bf 1916}, 688 (1916).
  %%CITATION = SPWPA,1916,688;%%
  %133 citations counted in INSPIRE as of 17 Nov 2017

  \bibitem{Einstein:1918btx}
  A.~Einstein,
  %``¨¹ber Gravitationswellen,''
  Sitzungsber.\ Preuss.\ Akad.\ Wiss.\ Berlin (Math.\ Phys.\ ) {\bf 1918}, 154 (1918).
  %%CITATION = SPWPA,1918,154;%%
  %123 citations counted in INSPIRE as of 17 Nov 2017

%\cite{Abbott:2016blz}
\bibitem{Abbott:2016blz}
  B.~P.~Abbott {\it et al.} [LIGO Scientific and Virgo Collaborations],
  %``Observation of Gravitational Waves from a Binary Black Hole Merger,''
  Phys.\ Rev.\ Lett.\  {\bf 116}, no. 6, 061102 (2016)
  doi:10.1103/PhysRevLett.116.061102
  [arXiv:1602.03837 [gr-qc]].
  %%CITATION = doi:10.1103/PhysRevLett.116.061102;%%
  %1993 citations counted in INSPIRE as of 07 Nov 2017

%\cite{TheLIGOScientific:2017qsa}
\bibitem{TheLIGOScientific:2017qsa}
  B.~P.~Abbott {\it et al.} [LIGO Scientific and Virgo Collaborations],
  %``GW170817: Observation of Gravitational Waves from a Binary Neutron Star Inspiral,''
  Phys.\ Rev.\ Lett.\  {\bf 119}, no. 16, 161101 (2017)
  doi:10.1103/PhysRevLett.119.161101
  [arXiv:1710.05832 [gr-qc]].
  %%CITATION = doi:10.1103/PhysRevLett.119.161101;%%
  %104 citations counted in INSPIRE as of 07 Nov 2017

\bibitem{Cai:2017cbj}
  R.~G.~Cai, Z.~Cao, Z.~K.~Guo, S.~J.~Wang and T.~Yang,
  %``The Gravitational-Wave Physics,''
  doi:10.1093/nsr/nwx029
  arXiv:1703.00187 [gr-qc].
  %%CITATION = doi:10.1093/nsr/nwx029;%%
  %16 citations counted in INSPIRE as of 18 Nov 2017

%\cite{Kosowsky:1992rz}
\bibitem{Kosowsky:1992rz}
  A.~Kosowsky, M.~S.~Turner and R.~Watkins,
  %``Gravitational waves from first order cosmological phase transitions,''
  Phys.\ Rev.\ Lett.\  {\bf 69}, 2026 (1992).
  doi:10.1103/PhysRevLett.69.2026
  %%CITATION = doi:10.1103/PhysRevLett.69.2026;%%
  %156 citations counted in INSPIRE as of 08 Nov 2017

%\cite{Kosowsky:1992vn}
\bibitem{Kosowsky:1992vn}
  A.~Kosowsky and M.~S.~Turner,
  %``Gravitational radiation from colliding vacuum bubbles: envelope approximation to many bubble collisions,''
  Phys.\ Rev.\ D {\bf 47}, 4372 (1993)
  doi:10.1103/PhysRevD.47.4372
  [astro-ph/9211004].
  %%CITATION = doi:10.1103/PhysRevD.47.4372;%%
  %140 citations counted in INSPIRE as of 08 Nov 2017

%\cite{Caprini:2015zlo}
\bibitem{Caprini:2015zlo}
  C.~Caprini {\it et al.},
  %``Science with the space-based interferometer eLISA. II: Gravitational waves from cosmological phase transitions,''
  JCAP {\bf 1604}, no. 04, 001 (2016)
  doi:10.1088/1475-7516/2016/04/001
  [arXiv:1512.06239 [astro-ph.CO]].
  %%CITATION = doi:10.1088/1475-7516/2016/04/001;%%
  %97 citations counted in INSPIRE as of 08 Nov 2017

%\cite{Jinno:2015doa}
\bibitem{Jinno:2015doa}
  R.~Jinno, K.~Nakayama and M.~Takimoto,
  %``Gravitational waves from the first order phase transition of the Higgs field at high energy scales,''
  Phys.\ Rev.\ D {\bf 93}, no. 4, 045024 (2016)
  doi:10.1103/PhysRevD.93.045024
  [arXiv:1510.02697 [hep-ph]].
  %%CITATION = doi:10.1103/PhysRevD.93.045024;%%
  %19 citations counted in INSPIRE as of 08 Nov 2017

%\cite{Kajantie:1996mn}
\bibitem{Kajantie:1996mn}
  K.~Kajantie, M.~Laine, K.~Rummukainen and M.~E.~Shaposhnikov,
  %``Is there a hot electroweak phase transition at m(H) larger or equal to m(W)?,''
  Phys.\ Rev.\ Lett.\  {\bf 77}, 2887 (1996)
  doi:10.1103/PhysRevLett.77.2887
  [hep-ph/9605288].
  %%CITATION = doi:10.1103/PhysRevLett.77.2887;%%
  %449 citations counted in INSPIRE as of 08 Nov 2017

%\cite{Gurtler:1997hr}
\bibitem{Gurtler:1997hr}
  M.~Gurtler, E.~M.~Ilgenfritz and A.~Schiller,
  %``Where the electroweak phase transition ends,''
  Phys.\ Rev.\ D {\bf 56}, 3888 (1997)
  doi:10.1103/PhysRevD.56.3888
  [hep-lat/9704013].
  %%CITATION = doi:10.1103/PhysRevD.56.3888;%%
  %135 citations counted in INSPIRE as of 08 Nov 2017

%\cite{Csikor:1998eu}
\bibitem{Csikor:1998eu}
  F.~Csikor, Z.~Fodor and J.~Heitger,
  %``Endpoint of the hot electroweak phase transition,''
  Phys.\ Rev.\ Lett.\  {\bf 82}, 21 (1999)
  doi:10.1103/PhysRevLett.82.21
  [hep-ph/9809291].
  %%CITATION = doi:10.1103/PhysRevLett.82.21;%%
  %207 citations counted in INSPIRE as of 08 Nov 2017

%\cite{Barger:2007im}
\bibitem{Barger:2007im}
  V.~Barger, P.~Langacker, M.~McCaskey, M.~J.~Ramsey-Musolf and G.~Shaughnessy,
  %``LHC Phenomenology of an Extended Standard Model with a Real Scalar Singlet,''
  Phys.\ Rev.\ D {\bf 77}, 035005 (2008)
  doi:10.1103/PhysRevD.77.035005
  [arXiv:0706.4311 [hep-ph]].
  %%CITATION = doi:10.1103/PhysRevD.77.035005;%%
  %355 citations counted in INSPIRE as of 08 Nov 2017

%\cite{Profumo:2007wc}
\bibitem{Profumo:2007wc}
  S.~Profumo, M.~J.~Ramsey-Musolf and G.~Shaughnessy,
  %``Singlet Higgs phenomenology and the electroweak phase transition,''
  JHEP {\bf 0708}, 010 (2007)
  doi:10.1088/1126-6708/2007/08/010
  [arXiv:0705.2425 [hep-ph]].
  %%CITATION = doi:10.1088/1126-6708/2007/08/010;%%
  %207 citations counted in INSPIRE as of 08 Nov 2017

%\cite{Damgaard:2015con}
\bibitem{Damgaard:2015con}
  P.~H.~Damgaard, A.~Haarr, D.~O'Connell and A.~Tranberg,
  %``Effective Field Theory and Electroweak Baryogenesis in the Singlet-Extended Standard Model,''
  JHEP {\bf 1602}, 107 (2016)
  doi:10.1007/JHEP02(2016)107
  [arXiv:1512.01963 [hep-ph]].
  %%CITATION = doi:10.1007/JHEP02(2016)107;%%
  %17 citations counted in INSPIRE as of 08 Nov 2017

%\cite{Vaskonen:2016yiu}
\bibitem{Vaskonen:2016yiu}
  V.~Vaskonen,
  %``Electroweak baryogenesis and gravitational waves from a real scalar singlet,''
  Phys.\ Rev.\ D {\bf 95}, no. 12, 123515 (2017)
  doi:10.1103/PhysRevD.95.123515
  [arXiv:1611.02073 [hep-ph]].
  %%CITATION = doi:10.1103/PhysRevD.95.123515;%%
  %23 citations counted in INSPIRE as of 08 Nov 2017

%\cite{Beniwal:2017eik}
\bibitem{Beniwal:2017eik}
  A.~Beniwal, M.~Lewicki, J.~D.~Wells, M.~White and A.~G.~Williams,
  %``Gravitational wave, collider and dark matter signals from a scalar singlet electroweak baryogenesis,''
  JHEP {\bf 1708}, 108 (2017)
  doi:10.1007/JHEP08(2017)108
  [arXiv:1702.06124 [hep-ph]].
  %%CITATION = doi:10.1007/JHEP08(2017)108;%%
  %19 citations counted in INSPIRE as of 08 Nov 2017

%\cite{Chen:2017qcz}
\bibitem{Chen:2017qcz}
  C.~Y.~Chen, J.~Kozaczuk and I.~M.~Lewis,
  %``Non-resonant Collider Signatures of a Singlet-Driven Electroweak Phase Transition,''
  JHEP {\bf 1708}, 096 (2017)
  doi:10.1007/JHEP08(2017)096
  [arXiv:1704.05844 [hep-ph]].
  %%CITATION = doi:10.1007/JHEP08(2017)096;%%
  %5 citations counted in INSPIRE as of 08 Nov 2017

%\cite{Cline:1996mga}
\bibitem{Cline:1996mga}
  J.~M.~Cline and P.~A.~Lemieux,
  %``Electroweak phase transition in two Higgs doublet models,''
  Phys.\ Rev.\ D {\bf 55}, 3873 (1997)
  doi:10.1103/PhysRevD.55.3873
  [hep-ph/9609240].
  %%CITATION = doi:10.1103/PhysRevD.55.3873;%%
  %94 citations counted in INSPIRE as of 08 Nov 2017

%\cite{Fromme:2006cm}
\bibitem{Fromme:2006cm}
  L.~Fromme, S.~J.~Huber and M.~Seniuch,
  %``Baryogenesis in the two-Higgs doublet model,''
  JHEP {\bf 0611}, 038 (2006)
  doi:10.1088/1126-6708/2006/11/038
  [hep-ph/0605242].
  %%CITATION = doi:10.1088/1126-6708/2006/11/038;%%
  %138 citations counted in INSPIRE as of 08 Nov 2017

%\cite{Dorsch:2013wja}
\bibitem{Dorsch:2013wja}
  G.~C.~Dorsch, S.~J.~Huber and J.~M.~No,
  %``A strong electroweak phase transition in the 2HDM after LHC8,''
  JHEP {\bf 1310}, 029 (2013)
  doi:10.1007/JHEP10(2013)029
  [arXiv:1305.6610 [hep-ph]].
  %%CITATION = doi:10.1007/JHEP10(2013)029;%%
  %59 citations counted in INSPIRE as of 08 Nov 2017

%\cite{Haarr:2016qzq}
\bibitem{Haarr:2016qzq}
  A.~Haarr, A.~Kvellestad and T.~C.~Petersen,
  %``Disfavouring Electroweak Baryogenesis and a hidden Higgs in a CP-violating Two-Higgs-Doublet Model,''
  arXiv:1611.05757 [hep-ph].
  %%CITATION = ARXIV:1611.05757;%%
  %8 citations counted in INSPIRE as of 08 Nov 2017

%\cite{Gunion:1989ci}
\bibitem{Gunion:1989ci}
  J.~F.~Gunion, R.~Vega and J.~Wudka,
  %``Higgs triplets in the standard model,''
  Phys.\ Rev.\ D {\bf 42}, 1673 (1990).
  doi:10.1103/PhysRevD.42.1673
  %%CITATION = doi:10.1103/PhysRevD.42.1673;%%
  %256 citations counted in INSPIRE as of 08 Nov 2017

%\cite{FileviezPerez:2008bj}
\bibitem{FileviezPerez:2008bj}
  P.~Fileviez Perez, H.~H.~Patel, M.~J.~Ramsey-Musolf and K.~Wang,
  %``Triplet Scalars and Dark Matter at the LHC,''
  Phys.\ Rev.\ D {\bf 79}, 055024 (2009)
  doi:10.1103/PhysRevD.79.055024
  [arXiv:0811.3957 [hep-ph]].
  %%CITATION = doi:10.1103/PhysRevD.79.055024;%%
  %80 citations counted in INSPIRE as of 08 Nov 2017

%\cite{Kuzmin:1985mm}
\bibitem{Kuzmin:1985mm}
  V.~A.~Kuzmin, V.~A.~Rubakov and M.~E.~Shaposhnikov,
  %``On the Anomalous Electroweak Baryon Number Nonconservation in the Early Universe,''
  Phys.\ Lett.\  {\bf 155B}, 36 (1985).
  doi:10.1016/0370-2693(85)91028-7
  %%CITATION = doi:10.1016/0370-2693(85)91028-7;%%
  %2440 citations counted in INSPIRE as of 08 Nov 2017

%\cite{Shaposhnikov:1987tw}
\bibitem{Shaposhnikov:1987tw}
  M.~E.~Shaposhnikov,
  %``Baryon Asymmetry of the Universe in Standard Electroweak Theory,''
  Nucl.\ Phys.\ B {\bf 287}, 757 (1987).
  doi:10.1016/0550-3213(87)90127-1
  %%CITATION = doi:10.1016/0550-3213(87)90127-1;%%
  %576 citations counted in INSPIRE as of 08 Nov 2017

%\cite{Weinberg:1975gm}
\bibitem{Weinberg:1975gm}
  S.~Weinberg,
  %``Implications of Dynamical Symmetry Breaking,''
  Phys.\ Rev.\ D {\bf 13}, 974 (1976)
  Addendum: [Phys.\ Rev.\ D {\bf 19}, 1277 (1979)].
  doi:10.1103/PhysRevD.19.1277, 10.1103/PhysRevD.13.974
  %%CITATION = doi:10.1103/PhysRevD.19.1277, 10.1103/PhysRevD.13.974;%%
  %2437 citations counted in INSPIRE as of 08 Nov 2017

%\cite{Susskind:1978ms}
\bibitem{Susskind:1978ms}
  L.~Susskind,
  %``Dynamics of Spontaneous Symmetry Breaking in the Weinberg-Salam Theory,''
  Phys.\ Rev.\ D {\bf 20}, 2619 (1979).
  doi:10.1103/PhysRevD.20.2619
  %%CITATION = doi:10.1103/PhysRevD.20.2619;%%
  %2486 citations counted in INSPIRE as of 08 Nov 2017
  
%\cite{Yamawaki:1985zg}
\bibitem{Yamawaki:1985zg}
  K.~Yamawaki, M.~Bando and K.~i.~Matumoto,
  %``Scale Invariant Technicolor Model and a Technidilaton,''
  Phys.\ Rev.\ Lett.\  {\bf 56}, 1335 (1986).
  doi:10.1103/PhysRevLett.56.1335
  %%CITATION = doi:10.1103/PhysRevLett.56.1335;%%
  %823 citations counted in INSPIRE as of 08 Nov 2017

%\cite{Bando:1986bg}
\bibitem{Bando:1986bg}
  M.~Bando, K.~i.~Matumoto and K.~Yamawaki,
  %``Technidilaton,''
  Phys.\ Lett.\ B {\bf 178}, 308 (1986).
  doi:10.1016/0370-2693(86)91516-9
  %%CITATION = doi:10.1016/0370-2693(86)91516-9;%%
  %170 citations counted in INSPIRE as of 08 Nov 2017

%\cite{Bando:1987we}
\bibitem{Bando:1987we}
  M.~Bando, T.~Morozumi, H.~So and K.~Yamawaki,
  %``Discriminating Technicolor Theories Through Flavor Changing Neutral Currents: Walking Or Standing Coupling Constants?,''
  Phys.\ Rev.\ Lett.\  {\bf 59}, 389 (1987).
  doi:10.1103/PhysRevLett.59.389
  %%CITATION = doi:10.1103/PhysRevLett.59.389;%%
  %140 citations counted in INSPIRE as of 08 Nov 2017

\bibitem{Cline:2008hr}
  J.~M.~Cline, M.~Jarvinen and F.~Sannino,
  %``The Electroweak Phase Transition in Nearly Conformal Technicolor,''
  Phys.\ Rev.\ D {\bf 78} (2008) 075027
  doi:10.1103/PhysRevD.78.075027
  [arXiv:0808.1512 [hep-ph]].
  %%CITATION = doi:10.1103/PhysRevD.78.075027;%%
  %35 citations counted in INSPIRE as of 09 Dec 2017

\bibitem{Jarvinen:2009mh}
  M.~Jarvinen, C.~Kouvaris and F.~Sannino,
  %``Gravitational Techniwaves,''
  Phys.\ Rev.\ D {\bf 81} (2010) 064027
  doi:10.1103/PhysRevD.81.064027
  [arXiv:0911.4096 [hep-ph]].
  %%CITATION = doi:10.1103/PhysRevD.81.064027;%%
  %13 citations counted in INSPIRE as of 09 Dec 2017

\bibitem{Fodor:2001au}
  Z.~Fodor and S.~D.~Katz,
  %``A New method to study lattice QCD at finite temperature and chemical potential,''
  Phys.\ Lett.\ B {\bf 534}, 87 (2002).

  \bibitem{Ding:2015ona}
  H.~T.~Ding, F.~Karsch and S.~Mukherjee,
  %``Thermodynamics of strong-interaction matter from Lattice QCD,''
  Int.\ J.\ Mod.\ Phys.\ E {\bf 24}, no. 10, 1530007 (2015)
  doi:10.1142/S0218301315300076
  [arXiv:1504.05274 [hep-lat]].

%\cite{Lucini:2012wq}
\bibitem{Lucini:2012wq}
  B.~Lucini, A.~Rago and E.~Rinaldi,
  %``SU($N_c$) gauge theories at deconfinement,''
  Phys.\ Lett.\ B {\bf 712}, 279 (2012)
  doi:10.1016/j.physletb.2012.04.070
  [arXiv:1202.6684 [hep-lat]].
  %%CITATION = doi:10.1016/j.physletb.2012.04.070;%%
  %30 citations counted in INSPIRE as of 08 Nov 2017

%\cite{Xu:2011pz}
\bibitem{Xu:2011pz}
  F.~Xu, H.~Mao, T.~K.~Mukherjee and M.~Huang,
  %``Dressed Polyakov loop and flavor dependent phase transitions,''
  Phys.\ Rev.\ D {\bf 84}, 074009 (2011)
  doi:10.1103/PhysRevD.84.074009
  [arXiv:1104.0873 [hep-ph]].
  %%CITATION = doi:10.1103/PhysRevD.84.074009;%%
  %12 citations counted in INSPIRE as of 19 Nov 2017

%\cite{Aharony:1999ti}
\bibitem{Aharony:1999ti}
  O.~Aharony, S.~S.~Gubser, J.~M.~Maldacena, H.~Ooguri and Y.~Oz,
  %``Large N field theories, string theory and gravity,''
  Phys.\ Rept.\  {\bf 323}, 183 (2000)
  doi:10.1016/S0370-1573(99)00083-6
  [hep-th/9905111].
  %%CITATION = doi:10.1016/S0370-1573(99)00083-6;%%
  %4149 citations counted in INSPIRE as of 09 Nov 2017

%\cite{Aharony:2002up}
\bibitem{Aharony:2002up}
  O.~Aharony,
  %``The NonAdS / nonCFT correspondence, or three different paths to QCD,''
  hep-th/0212193.
  %%CITATION = HEP-TH/0212193;%%
  %70 citations counted in INSPIRE as of 09 Nov 2017

%\cite{Zaffaroni:2005ty}
\bibitem{Zaffaroni:2005ty}
  A.~Zaffaroni,
  %``RTN lectures on the non AdS / non CFT correspondence,''
  PoS RTN {\bf 2005}, 005 (2005).
  %%CITATION = POSCI,RTN2005,005;%%
  %18 citations counted in INSPIRE as of 09 Nov 2017

%\cite{Erdmenger:2007cm}
\bibitem{Erdmenger:2007cm}
  J.~Erdmenger, N.~Evans, I.~Kirsch and E.~Threlfall,
  %``Mesons in Gauge/Gravity Duals - A Review,''
  Eur.\ Phys.\ J.\ A {\bf 35}, 81 (2008)
  doi:10.1140/epja/i2007-10540-1
  [arXiv:0711.4467 [hep-th]].
  %%CITATION = doi:10.1140/epja/i2007-10540-1;%%
  %376 citations counted in INSPIRE as of 09 Nov 2017

%\cite{Maldacena:1997re}
\bibitem{Maldacena:1997re}
  J.~M.~Maldacena,
  %``The Large N limit of superconformal field theories and supergravity,''
  Int.\ J.\ Theor.\ Phys.\  {\bf 38}, 1113 (1999)
  [Adv.\ Theor.\ Math.\ Phys.\  {\bf 2}, 231 (1998)]
  doi:10.1023/A:1026654312961
  [hep-th/9711200].
  %%CITATION = doi:10.1023/A:1026654312961;%%
  %13194 citations counted in INSPIRE as of 08 Nov 2017

%\cite{Gubser:1998bc}
\bibitem{Gubser:1998bc}
  S.~S.~Gubser, I.~R.~Klebanov and A.~M.~Polyakov,
  %``Gauge theory correlators from noncritical string theory,''
  Phys.\ Lett.\ B {\bf 428}, 105 (1998)
  doi:10.1016/S0370-2693(98)00377-3
  [hep-th/9802109].
  %%CITATION = doi:10.1016/S0370-2693(98)00377-3;%%
  %7426 citations counted in INSPIRE as of 08 Nov 2017

%\cite{Witten:1998qj}
\bibitem{Witten:1998qj}
  E.~Witten,
  %``Anti-de Sitter space and holography,''
  Adv.\ Theor.\ Math.\ Phys.\  {\bf 2}, 253 (1998)
  [hep-th/9802150].
  %%CITATION = HEP-TH/9802150;%%
  %8640 citations counted in INSPIRE as of 08 Nov 2017

%\cite{Erlich:2005qh}
\bibitem{Erlich:2005qh}
  J.~Erlich, E.~Katz, D.~T.~Son and M.~A.~Stephanov,
  %``QCD and a holographic model of hadrons,''
  Phys.\ Rev.\ Lett.\  {\bf 95}, 261602 (2005)
  doi:10.1103/PhysRevLett.95.261602
  [hep-ph/0501128].
  %%CITATION = doi:10.1103/PhysRevLett.95.261602;%%
  %816 citations counted in INSPIRE as of 09 Nov 2017

%\cite{DaRold:2005vr}
\bibitem{DaRold:2005vr}
  L.~Da Rold and A.~Pomarol,
  %``The Scalar and pseudoscalar sector in a five-dimensional approach to chiral symmetry breaking,''
  JHEP {\bf 0601}, 157 (2006)
  doi:10.1088/1126-6708/2006/01/157
  [hep-ph/0510268].
  %%CITATION = doi:10.1088/1126-6708/2006/01/157;%%
  %135 citations counted in INSPIRE as of 09 Nov 2017

%\cite{Karch:2006pv}
\bibitem{Karch:2006pv}
  A.~Karch, E.~Katz, D.~T.~Son and M.~A.~Stephanov,
  %``Linear confinement and AdS/QCD,''
  Phys.\ Rev.\ D {\bf 74}, 015005 (2006)
  doi:10.1103/PhysRevD.74.015005
  [hep-ph/0602229].
  %%CITATION = doi:10.1103/PhysRevD.74.015005;%%
  %747 citations counted in INSPIRE as of 09 Nov 2017

\bibitem{mesons}
T.~Sakai and S.~Sugimoto,
%``Low energy hadron physics in holographic QCD,''
Prog.\ Theor.\ Phys.\ \textbf{113}, 843 (2005); %  [arXiv:hep-th/0412141].
%%CITATION = PTPKA,113,843;%%
% T.~Sakai and S.~Sugimoto,
%``More on a holographic dual of QCD,''
Prog.\ Theor.\ Phys.\ \textbf{114}, 1083 (2006); % [arXiv:hep-th/0507073].
%%CITATION = PTPKA,114,1083;%%
G.~F.~de Teramond and S.~J.~Brodsky,
%``The hadronic spectrum of a holographic dual of QCD,''
Phys.\ Rev.\ Lett.\ \textbf{94}, 201601 (2005);%  [arXiv:hep-th/0501022].
%%CITATION = PRLTA,94,201601;%%
%\bibitem{DaRold2005}
L.~Da Rold and A.~Pomarol,
%``Chiral symmetry breaking from five dimensional spaces,''
Nucl.\ Phys.\ B \textbf{721}, 79 (2005);% [arXiv:hep-ph/0501218].
%%CITATION = NUPHA,B721,79;%%
%\bibitem{Ghoroku:2005vt}
K.~Ghoroku, N.~Maru, M.~Tachibana and M.~Yahiro,
%``Holographic model for hadrons in deformed AdS(5) background,''
Phys.\ Lett.\ B \textbf{633}, 602 (2006);
% [arXiv:hep-ph/0510334].
%%CITATION = PHLTA,B633,602;%%
%\bibitem{and06}
O. Andreev, V.I. Zakharov, arXiv:hep-ph/0703010; Phys. Rev.
D \textbf{74}, 025023 \ (2006);
%\bibitem{kru05}
M. Kruczenski, L. A. P. Zayas, J. Sonnenschein and D. Vaman,
JHEP \textbf{06}, 046 (2005);
S. Kuperstein and J. Sonnenschein, JHEP\textbf{11}, 026 (2004).
%\bibitem{Forkel}
 H.~Forkel, M.~Beyer and T.~Frederico,
%``Linear square-mass trajectories of radially and orbitally excited hadrons
%in holographic QCD,''
JHEP \textbf{0707}, 077 (2007); %  [arXiv:0705.1857 [hep-ph]].
%%CITATION = JHEPA,0707,077;%%
%\cite{Chen:2015zhh}
%\bibitem{Chen:2015zhh}
  Y.~Chen and M.~Huang,
  %``Two-gluon and trigluon glueballs from dynamical holography QCD,''
  Chin.\ Phys.\ C {\bf 40}, no. 12, 123101 (2016)
  %doi:10.1088/1674-1137/40/12/123101
  %[arXiv:1511.07018 [hep-ph]].
  %%CITATION = doi:10.1088/1674-1137/40/12/123101;%%
  %8 citations counted in INSPIRE as of 09 Nov 2017

\bibitem{bns}
%\cite{DeWolfe:2010he}
%\bibitem{DeWolfe:2010he}
  O.~DeWolfe, S.~S.~Gubser and C.~Rosen,
  %``A holographic critical point,''
  Phys.\ Rev.\ D {\bf 83}, 086005 (2011);
  %doi:10.1103/PhysRevD.83.086005
  %[arXiv:1012.1864 [hep-th]].
  %%CITATION = doi:10.1103/PhysRevD.83.086005;%%
  %58 citations counted in INSPIRE as of 09 Nov 2017
%\cite{DeWolfe:2011ts}
%\bibitem{DeWolfe:2011ts}
  O.~DeWolfe, S.~S.~Gubser and C.~Rosen,
  %``Dynamic critical phenomena at a holographic critical point,''
  Phys.\ Rev.\ D {\bf 84}, 126014 (2011);
  %doi:10.1103/PhysRevD.84.126014
  %[arXiv:1108.2029 [hep-th]].
  %%CITATION = doi:10.1103/PhysRevD.84.126014;%%
  %42 citations counted in INSPIRE as of 09 Nov 2017
%\cite{Yang:2014bqa}
%\bibitem{Yang:2014bqa}
  Y.~Yang and P.~H.~Yuan,
  %``A Refined Holographic QCD Model and QCD Phase Structure,''
  JHEP {\bf 1411}, 149 (2014);
  %doi:10.1007/JHEP11(2014)149
  %[arXiv:1406.1865 [hep-th]].
  %%CITATION = doi:10.1007/JHEP11(2014)149;%%
  %6 citations counted in INSPIRE as of 09 Nov 2017
%\cite{Critelli:2017oub}
%\bibitem{Critelli:2017oub}
  R.~Critelli, J.~Noronha, J.~Noronha-Hostler, I.~Portillo, C.~Ratti and R.~Rougemont,
  %``Black-hole engineered critical point in the phase diagram of primordial quark-gluon matter,''
  arXiv:1706.00455 [nucl-th];
  %%CITATION = ARXIV:1706.00455;%%
  %7 citations counted in INSPIRE as of 09 Nov 2017
%\cite{Li:2017ple}
%\bibitem{Li:2017ple}
  Z.~Li, Y.~Chen, D.~Li and M.~Huang,
  %``Locating the QCD critical end point through the peaked baryon number susceptibilities along the freeze-out line,''
  arXiv:1706.02238 [hep-ph].
  %%CITATION = ARXIV:1706.02238;%%
  %2 citations counted in INSPIRE as of 09 Nov 2017

%\cite{Haba:2010hu}
\bibitem{Haba:2010hu}
  K.~Haba, S.~Matsuzaki and K.~Yamawaki,
  %``Holographic Techni-dilaton,''
  Phys.\ Rev.\ D {\bf 82}, 055007 (2010)
  doi:10.1103/PhysRevD.82.055007
  [arXiv:1006.2526 [hep-ph]].
  %%CITATION = doi:10.1103/PhysRevD.82.055007;%%
  %49 citations counted in INSPIRE as of 09 Nov 2017

%\cite{Matsuzaki:2012xx}
\bibitem{Matsuzaki:2012xx}
  S.~Matsuzaki and K.~Yamawaki,
  %``Holographic techni-dilaton at 125 GeV,''
  Phys.\ Rev.\ D {\bf 86}, 115004 (2012)
  doi:10.1103/PhysRevD.86.115004
  [arXiv:1209.2017 [hep-ph]].
  %%CITATION = doi:10.1103/PhysRevD.86.115004;%%
  %76 citations counted in INSPIRE as of 09 Nov 2017

\bibitem{htc}
%\cite{Hong:2006si}
%\bibitem{Hong:2006si}
  D.~K.~Hong and H.~U.~Yee,
  %``Holographic estimate of oblique corrections for technicolor,''
  Phys.\ Rev.\ D {\bf 74}, 015011 (2006);
  %doi:10.1103/PhysRevD.74.015011
  %[hep-ph/0602177].
  %%CITATION = doi:10.1103/PhysRevD.74.015011;%%
  %102 citations counted in INSPIRE as of 09 Nov 2017
%\cite{Hirn:2006nt}
%\bibitem{Hirn:2006nt}
  J.~Hirn and V.~Sanz,
  %``A Negative S parameter from holographic technicolor,''
  Phys.\ Rev.\ Lett.\  {\bf 97}, 121803 (2006);
  %doi:10.1103/PhysRevLett.97.121803
  %[hep-ph/0606086].
  %%CITATION = doi:10.1103/PhysRevLett.97.121803;%%
  %116 citations counted in INSPIRE as of 09 Nov 2017
%\cite{Piai:2006hy}
%\bibitem{Piai:2006hy}
  M.~Piai,
  %``Precision electro-weak parameters from AdS(5), localized kinetic terms and anomalous dimensions,''
  hep-ph/0608241;
  %%CITATION = HEP-PH/0608241;%%
  %52 citations counted in INSPIRE as of 09 Nov 2017
%\cite{Carone:2006wj}
%\bibitem{Carone:2006wj}
  C.~D.~Carone, J.~Erlich and J.~A.~Tan,
  %``Holographic Bosonic Technicolor,''
  Phys.\ Rev.\ D {\bf 75}, 075005 (2007);
  %doi:10.1103/PhysRevD.75.075005
  %[hep-ph/0612242].
  %%CITATION = doi:10.1103/PhysRevD.75.075005;%%
  %70 citations counted in INSPIRE as of 09 Nov 2017
%\cite{Nunez:2008wi}
%\bibitem{Nunez:2008wi}
  C.~Nunez, I.~Papadimitriou and M.~Piai,
  %``Walking Dynamics from String Duals,''
  Int.\ J.\ Mod.\ Phys.\ A {\bf 25}, 2837 (2010);
  %doi:10.1142/S0217751X10049189
  %[arXiv:0812.3655 [hep-th]].
  %%CITATION = doi:10.1142/S0217751X10049189;%%
  %78 citations counted in INSPIRE as of 09 Nov 2017
%\cite{Anguelova:2011bc}
%\bibitem{Anguelova:2011bc}
  L.~Anguelova, P.~Suranyi and L.~C.~R.~Wijewardhana,
  %``Holographic Walking Technicolor from D-branes,''
  Nucl.\ Phys.\ B {\bf 852}, 39 (2011);
  %doi:10.1016/j.nuclphysb.2011.06.010
  %[arXiv:1105.4185 [hep-th]].
  %%CITATION = doi:10.1016/j.nuclphysb.2011.06.010;%%
  %31 citations counted in INSPIRE as of 09 Nov 2017
%\cite{Anguelova:2012ka}
%\bibitem{Anguelova:2012ka}
  L.~Anguelova, P.~Suranyi and L.~C.~R.~Wijewardhana,
  %``Scalar Mesons in Holographic Walking Technicolor,''
  Nucl.\ Phys.\ B {\bf 862}, 671 (2012);
  %doi:10.1016/j.nuclphysb.2012.05.005
  %[arXiv:1203.1968 [hep-th]].
  %%CITATION = doi:10.1016/j.nuclphysb.2012.05.005;%%
  %26 citations counted in INSPIRE as of 09 Nov 2017
%\cite{Elander:2012fk}
%\bibitem{Elander:2012fk}
  D.~Elander and M.~Piai,
  %``The decay constant of the holographic techni-dilaton and the 125 GeV boson,''
  Nucl.\ Phys.\ B {\bf 867}, 779 (2013);
  %doi:10.1016/j.nuclphysb.2012.10.019
  %[arXiv:1208.0546 [hep-ph]].
  %%CITATION = doi:10.1016/j.nuclphysb.2012.10.019;%%
  %66 citations counted in INSPIRE as of 09 Nov 2017

\bibitem{hchm}
%\cite{Contino:2003ve}
%\bibitem{Contino:2003ve}
  R.~Contino, Y.~Nomura and A.~Pomarol,
  %``Higgs as a holographic pseudoGoldstone boson,''
  Nucl.\ Phys.\ B {\bf 671}, 148 (2003);
  %doi:10.1016/j.nuclphysb.2003.08.027
  %[hep-ph/0306259].
  %%CITATION = doi:10.1016/j.nuclphysb.2003.08.027;%%
  %504 citations counted in INSPIRE as of 09 Nov 2017
%\cite{Agashe:2004rs}
%\bibitem{Agashe:2004rs}
  K.~Agashe, R.~Contino and A.~Pomarol,
  %``The Minimal composite Higgs model,''
  Nucl.\ Phys.\ B {\bf 719}, 165 (2005);
  %doi:10.1016/j.nuclphysb.2005.04.035
  %[hep-ph/0412089].
  %%CITATION = doi:10.1016/j.nuclphysb.2005.04.035;%%
  %1016 citations counted in INSPIRE as of 09 Nov 2017
%\cite{Agashe:2005dk}
%\bibitem{Agashe:2005dk}
  K.~Agashe and R.~Contino,
  %``The Minimal composite Higgs model and electroweak precision tests,''
  Nucl.\ Phys.\ B {\bf 742}, 59 (2006);
  %doi:10.1016/j.nuclphysb.2006.02.011
  %[hep-ph/0510164].
  %%CITATION = doi:10.1016/j.nuclphysb.2006.02.011;%%
  %201 citations counted in INSPIRE as of 09 Nov 2017
%\cite{Croon:2015wba}
%\bibitem{Croon:2015wba}
  D.~Croon, B.~M.~Dillon, S.~J.~Huber and V.~Sanz,
  %``Exploring holographic Composite Higgs models,''
  JHEP {\bf 1607}, 072 (2016);
  %doi:10.1007/JHEP07(2016)072
  %[arXiv:1510.08482 [hep-ph]].
  %%CITATION = doi:10.1007/JHEP07(2016)072;%%
  %7 citations counted in INSPIRE as of 09 Nov 2017
%\cite{Espriu:2017mlq}
%\bibitem{Espriu:2017mlq}
  D.~Espriu and A.~Katanaeva,
  %``Holographic description of $SO(5) \rightarrow SO(4)$ composite Higgs model,''
  arXiv:1706.02651 [hep-ph].
  %%CITATION = ARXIV:1706.02651;%%

\bibitem{Adams:2012th}
  A.~Adams, L.~D.~Carr, T.~Schaefer, P.~Steinberg and J.~E.~Thomas,
  ``Strongly Correlated Quantum Fluids: Ultracold Quantum Gases, Quantum Chromodynamic Plasmas, and Holographic Duality,''  New J.\ Phys.\  {\bf 14}, 115009 (2012)  [arXiv:1205.5180 [hep-th]].  %%CITATION = ARXIV:1205.5180;%%

\bibitem{Andreev:2006ct}
  O.~Andreev and V.~I.~Zakharov,
 % ``Heavy-quark potentials and AdS/QCD,''
  Phys.\ Rev.\  D {\bf 74}, 025023 (2006).
  %%[arXiv:hep-ph/0604204].
  %%CITATION = PHRVA,D74,025023;%%

\bibitem{Pirner:2009gr}
  H.~J.~Pirner and B.~Galow,
  %``Strong Equivalence of the AdS-Metric and the QCD Running Coupling,''
  Phys.\ Lett.\  B {\bf 679}, 51 (2009).
 % [arXiv:0903.2701 [hep-ph]].
  %%CITATION = PHLTA,B679,51;%%
  %\cite{He:2010ye}

\bibitem{EKSS2005}
J.~Erlich, E.~Katz, D.~T.~Son and M.~A.~Stephanov,
%``QCD and a holographic model of hadrons,''
Phys.\ Rev.\ Lett.\ \textbf{95}, 261602 (2005); % [arXiv:hep-ph/0501128].
%%CITATION = PRLTA,95,261602;%%

\bibitem{Colangelo:2008us}
P.~Colangelo, F.~De Fazio, F.~Giannuzzi, F.~Jugeau and S.~Nicotri,
 ``Light scalar mesons in the soft-wall model of AdS/QCD,''  Phys.\ Rev.\ D {\bf 78}, 055009 (2008)  [arXiv:0807.1054 [hep-ph]].  %%CITATION = ARXIV:0807.1054;%%  %86 citations counted in INSPIRE as of 14 Mar 2013

\bibitem{Gherghetta-Kapusta-Kelley}
T.~Gherghetta, J.~I.~Kapusta and T.~M.~Kelley,
%  ``Chiral symmetry breaking in the soft-wall AdS/QCD model,''
Phys.\ Rev.\ D {\bf 79} (2009) 076003; % [arXiv:0902.1998 [hep-ph]].  %%CITATION = ARXIV:0902.1998;%%

\bibitem{YLWu}
  Y.~-Q.~Sui, Y.~-L.~Wu, Z.~-F.~Xie and Y.~-B.~Yang,
  %``Prediction for the Mass Spectra of Resonance Mesons in the Soft-Wall AdS/QCD with a Modified 5D Metric,''
  Phys.\ Rev.\ D {\bf 81} (2010) 014024;  % [arXiv:0909.3887 [hep-ph]].  %%CITATION = ARXIV:0909.3887;%%
Y.~-Q.~Sui, Y.~-L.~Wu and Y.~-B.~Yang,
  %``Predictive AdS/QCD Model for Mass Spectra of Mesons with Three Flavors,''
  Phys.\ Rev.\ D {\bf 83} (2011) 065030.  %[arXiv:1012.3518 [hep-ph]].  %%CITATION = ARXIV:1012.3518;%%


\bibitem{Li:2012ay}
  D.~Li, M.~Huang and Q.~S.~Yan,
  %``A dynamical soft-wall holographic QCD model for chiral symmetry breaking and linear confinement,''
  Eur.\ Phys.\ J.\ C {\bf 73}, 2615 (2013)
  doi:10.1140/epjc/s10052-013-2615-3
  [arXiv:1206.2824 [hep-th]].
  %%CITATION = doi:10.1140/epjc/s10052-013-2615-3;%%
  %27 citations counted in INSPIRE as of 28 Nov 2017

  \bibitem{Li:2013oda}
  D.~Li and M.~Huang,
  %``Dynamical holographic QCD model for glueball and light meson spectra,''
  JHEP {\bf 1311}, 088 (2013)
  doi:10.1007/JHEP11(2013)088
  [arXiv:1303.6929 [hep-ph]].
  %%CITATION = doi:10.1007/JHEP11(2013)088;%%
  %35 citations counted in INSPIRE as of 28 Nov 2017

\bibitem{Li:2011hp}
  D.~Li, S.~He, M.~Huang and Q.~S.~Yan,
  %``Thermodynamics of deformed AdS$_5$ model with a positive/negative quadratic correction in graviton-dilaton system,''
  JHEP {\bf 1109}, 041 (2011)
  doi:10.1007/JHEP09(2011)041
  [arXiv:1103.5389 [hep-th]].
  %%CITATION = doi:10.1007/JHEP09(2011)041;%%
  %43 citations counted in INSPIRE as of 28 Nov 2017

\bibitem{Li:2014hja}
  D.~Li, J.~Liao and M.~Huang,
  %``Enhancement of jet quenching around phase transition: result from the dynamical holographic model,''
  Phys.\ Rev.\ D {\bf 89}, no. 12, 126006 (2014)
  doi:10.1103/PhysRevD.89.126006
  [arXiv:1401.2035 [hep-ph]].
  %%CITATION = doi:10.1103/PhysRevD.89.126006;%%
  %28 citations counted in INSPIRE as of 09 Dec 2017

\bibitem{Li:2014dsa}
  D.~Li, S.~He and M.~Huang,
  %``Temperature dependent transport coefficients in a dynamical holographic QCD model,''
  JHEP {\bf 1506}, 046 (2015)
  doi:10.1007/JHEP06(2015)046
  [arXiv:1411.5332 [hep-ph]].
  %%CITATION = doi:10.1007/JHEP06(2015)046;%%
  %17 citations counted in INSPIRE as of 28 Nov 2017

  \bibitem{Chen:2015zhh}
  Y.~Chen and M.~Huang,
  %``Two-gluon and trigluon glueballs from dynamical holography QCD,''
  Chin.\ Phys.\ C {\bf 40}, no. 12, 123101 (2016)
  doi:10.1088/1674-1137/40/12/123101
  [arXiv:1511.07018 [hep-ph]].
  %%CITATION = doi:10.1088/1674-1137/40/12/123101;%%
  %8 citations counted in INSPIRE as of 28 Nov 2017
%\cite{Patrignani:2016xqp}
\bibitem{Patrignani:2016xqp}
  C.~Patrignani {\it et al.} [Particle Data Group],
  %``Review of Particle Physics,''
  Chin.\ Phys.\ C {\bf 40}, no. 10, 100001 (2016).
  doi:10.1088/1674-1137/40/10/100001
  %%CITATION = doi:10.1088/1674-1137/40/10/100001;%%
  %1879 citations counted in INSPIRE as of 09 Nov 2017

%\cite{Huber:2008hg}
\bibitem{Huber:2008hg}
  S.~J.~Huber and T.~Konstandin,
  %``Gravitational Wave Production by Collisions: More Bubbles,''
  JCAP {\bf 0809}, 022 (2008)
  doi:10.1088/1475-7516/2008/09/022
  [arXiv:0806.1828 [hep-ph]].
  %%CITATION = doi:10.1088/1475-7516/2008/09/022;%%
  %137 citations counted in INSPIRE as of 09 Nov 2017

%\cite{Jinno:2016vai}
\bibitem{Jinno:2016vai}
  R.~Jinno and M.~Takimoto,
  %``Gravitational waves from bubble collisions: An analytic derivation,''
  Phys.\ Rev.\ D {\bf 95}, no. 2, 024009 (2017)
  doi:10.1103/PhysRevD.95.024009
  [arXiv:1605.01403 [astro-ph.CO]].
  %%CITATION = doi:10.1103/PhysRevD.95.024009;%%
  %24 citations counted in INSPIRE as of 09 Nov 2017

%\cite{Hindmarsh:2015qta}
\bibitem{Hindmarsh:2015qta}
  M.~Hindmarsh, S.~J.~Huber, K.~Rummukainen and D.~J.~Weir,
  %``Numerical simulations of acoustically generated gravitational waves at a first order phase transition,''
  Phys.\ Rev.\ D {\bf 92}, no. 12, 123009 (2015)
  doi:10.1103/PhysRevD.92.123009
  [arXiv:1504.03291 [astro-ph.CO]].
  %%CITATION = doi:10.1103/PhysRevD.92.123009;%%
  %58 citations counted in INSPIRE as of 09 Nov 2017

%\cite{Hindmarsh:2017gnf}
\bibitem{Hindmarsh:2017gnf}
  M.~Hindmarsh, S.~J.~Huber, K.~Rummukainen and D.~J.~Weir,
  %``Shape of the acoustic gravitational wave power spectrum from a first order phase transition,''
  arXiv:1704.05871 [astro-ph.CO].
  %%CITATION = ARXIV:1704.05871;%%
  %6 citations counted in INSPIRE as of 09 Nov 2017

%\cite{Caprini:2009yp}
\bibitem{Caprini:2009yp}
  C.~Caprini, R.~Durrer and G.~Servant,
  %``The stochastic gravitational wave background from turbulence and magnetic fields generated by a first-order phase transition,''
  JCAP {\bf 0912}, 024 (2009)
  doi:10.1088/1475-7516/2009/12/024
  [arXiv:0909.0622 [astro-ph.CO]].
  %%CITATION = doi:10.1088/1475-7516/2009/12/024;%%
  %107 citations counted in INSPIRE as of 09 Nov 2017

%\cite{Binetruy:2012ze}
\bibitem{Binetruy:2012ze}
  P.~Binetruy, A.~Bohe, C.~Caprini and J.~F.~Dufaux,
  %``Cosmological Backgrounds of Gravitational Waves and eLISA/NGO: Phase Transitions, Cosmic Strings and Other Sources,''
  JCAP {\bf 1206}, 027 (2012)
  doi:10.1088/1475-7516/2012/06/027
  [arXiv:1201.0983 [gr-qc]].
  %%CITATION = doi:10.1088/1475-7516/2012/06/027;%%
  %105 citations counted in INSPIRE as of 09 Nov 2017

%\cite{Weir:2017wfa}
\bibitem{Weir:2017wfa}
  D.~J.~Weir,
  %``Gravitational waves from a first order electroweak phase transition: a review,''
  arXiv:1705.01783 [hep-ph].
  %%CITATION = ARXIV:1705.01783;%%
  %2 citations counted in INSPIRE as of 09 Nov 2017

%\cite{Steinhardt:1981ct}
\bibitem{Steinhardt:1981ct}
  P.~J.~Steinhardt,
  %``Relativistic Detonation Waves and Bubble Growth in False Vacuum Decay,''
  Phys.\ Rev.\ D {\bf 25}, 2074 (1982).
  doi:10.1103/PhysRevD.25.2074
  %%CITATION = doi:10.1103/PhysRevD.25.2074;%%
  %128 citations counted in INSPIRE as of 09 Nov 2017

%\cite{Kamionkowski:1993fg}
\bibitem{Kamionkowski:1993fg}
  M.~Kamionkowski, A.~Kosowsky and M.~S.~Turner,
  %``Gravitational radiation from first order phase transitions,''
  Phys.\ Rev.\ D {\bf 49}, 2837 (1994)
  doi:10.1103/PhysRevD.49.2837
  [astro-ph/9310044].
  %%CITATION = doi:10.1103/PhysRevD.49.2837;%%
  %280 citations counted in INSPIRE as of 09 Nov 2017

%\cite{Nicolis:2003tg}
\bibitem{Nicolis:2003tg}
  A.~Nicolis,
  %``Relic gravitational waves from colliding bubbles and cosmic turbulence,''
  Class.\ Quant.\ Grav.\  {\bf 21}, L27 (2004)
  doi:10.1088/0264-9381/21/4/L05
  [gr-qc/0303084].
  %%CITATION = doi:10.1088/0264-9381/21/4/L05;%%
  %76 citations counted in INSPIRE as of 09 Nov 2017

%\cite{Espinosa:2010hh}
\bibitem{Espinosa:2010hh}
  J.~R.~Espinosa, T.~Konstandin, J.~M.~No and G.~Servant,
  %``Energy Budget of Cosmological First-order Phase Transitions,''
  JCAP {\bf 1006}, 028 (2010)
  doi:10.1088/1475-7516/2010/06/028
  [arXiv:1004.4187 [hep-ph]].
  %%CITATION = doi:10.1088/1475-7516/2010/06/028;%%
  %92 citations counted in INSPIRE as of 09 Nov 2017

%\cite{Ahmadvand:2017xrw}
\bibitem{Ahmadvand:2017xrw}
  M.~Ahmadvand and K.~Bitaghsir Fadafan,
  %``Gravitational waves generated from the cosmological QCD phase transition within AdS/QCD,''
  Phys.\ Lett.\ B {\bf 772}, 747 (2017)
  doi:10.1016/j.physletb.2017.07.039
  [arXiv:1703.02801 [hep-th]].
  %%CITATION = doi:10.1016/j.physletb.2017.07.039;%%
  %2 citations counted in INSPIRE as of 09 Nov 2017

%\cite{Ahmadvand:2017tue}
\bibitem{Ahmadvand:2017tue}
  M.~Ahmadvand and K.~Bitaghsir Fadafan,
  %``The cosmic QCD phase transition with dense matter and its gravitational waves from holography,''
  arXiv:1707.05068 [hep-th].
  %%CITATION = ARXIV:1707.05068;%%
  %1 citations counted in INSPIRE as of 09 Nov 2017

\bibitem{GW-detector}
http://rhcole.com/apps/GWplotter/

\bibitem{Kuroda:2015owv}
  K.~Kuroda, W.~T.~Ni and W.~P.~Pan,
  %``Gravitational waves: Classification, Methods of detection, Sensitivities, and Sources,''
  Int.\ J.\ Mod.\ Phys.\ D {\bf 24}, no. 14, 1530031 (2015)
  doi:10.1142/S0218271815300311
  [arXiv:1511.00231 [gr-qc]].
  %%CITATION = doi:10.1142/S0218271815300311;%%
  %16 citations counted in INSPIRE as of 09 Dec 2017
  
\bibitem{FAST}
 R.~Nan {\it et al.},
  %``The Five-Hundred-Meter Aperture Spherical Radio Telescope (FAST) Project,''
  Int.\ J.\ Mod.\ Phys.\ D {\bf 20}, 989 (2011)
  doi:10.1142/S0218271811019335
  [arXiv:1105.3794 [astro-ph.IM]].
  %%CITATION = doi:10.1142/S0218271811019335;%%
  %80 citations counted in INSPIRE as of 09 Dec 2017
   
\end{thebibliography}
\end{document}